\documentclass[aps, pra, superscriptaddress , amsmath, amssymb, twocolumn, nofootinbib]{revtex4-1}

\usepackage[T1]{fontenc}
\usepackage{amsmath,amsfonts,bbm, graphicx, color}
\usepackage{perpage}
\MakePerPage{footnote}
\usepackage{amssymb}
\usepackage{appendix}
\usepackage[colorlinks]{hyperref}
\usepackage[figure,table]{hypcap}
\usepackage[textsize=tiny]{todonotes}
\usepackage{mathtools}

\hypersetup{
	bookmarksnumbered,
	pdfstartview={FitH},
	citecolor={green},
	linkcolor={darkblue},
	urlcolor={darkblue},
	pdfpagemode={UseOutlines}}
\definecolor{darkgreen}{RGB}{20,100,20}
\definecolor{darkblue}{RGB}{0,0,130}
\definecolor{darkred}{rgb}{.8,0,0}

\def\>{\rangle}
\def\<{\langle}

\def\>{\rangle}
\def\<{\langle}
\newcommand{\ket}[1]{|#1 \rangle}
\newcommand{\bra}[1]{\langle #1|}

\newcommand{\bi[1]}{\hat{b}_{\text{I}#1}}
\newcommand{\bl[1]}{\hat{b}_{\Lambda#1}}
\newcommand{\bii[1]}{\hat{b}_{\text{II}#1}}
\newcommand{\wi[1]}{w_{\text{I}#1}}
\newcommand{\wii[1]}{w_{\text{II}#1}}
\newcommand{\psii}{\psi_\text{I}}
\newcommand{\psiii}{\psi_\text{II}}
\newcommand{\phii}{\phi_\text{I}}
\newcommand{\phiii}{\phi_\text{II}}
\newcommand{\ali}{\alpha_\text{I}}
\newcommand{\alii}{\alpha_\text{II}}
\newcommand{\bei}{\beta_\text{I}}
\newcommand{\beii}{\beta_\text{II}}

\def\d{^{\dag}}
\renewcommand{\Re}{\text{Re}} 
\renewcommand{\Im}{\text{Im}}

\def\I{ \mathbbm{1} }

\def\>{\rangle}
\def\<{\langle}
\def \be{\begin{equation}}
\def \ee{\end{equation}}
\def \beq{\begin{equation}}
\def \eeq{\end{equation}}
\def \bea{\begin{eqnarray}}
\def \eea{\end{eqnarray}}

\renewcommand{\[}{\begin{equation}}
\renewcommand{\]}{\end{equation}}

\newcommand{\abs}[1]{|#1|}

\definecolor{darkred}{rgb}{.8,0,0}
\definecolor{magenta}{RGB}{255,0,255}
\definecolor{green}{rgb}{.2,.6,.2}

\usepackage{soul}

\begin{document}

\title{Effect of  relativistic acceleration on localized two-mode Gaussian quantum states}


\author{Mehdi Ahmadi}
\email{mehdi.ahmadi@fuw.edu.pl}
\affiliation{Institute of Theoretical Physics, University of Warsaw, Pasteura 5, 02-093 Warsaw, Poland}
\author{Krzysztof Lorek}
\email{krzysztof.lorek@fuw.edu.pl}
\affiliation{Institute of Theoretical Physics, University of Warsaw, Pasteura 5, 02-093 Warsaw, Poland}
\author{Agata Ch{\k{e}}ci\'nska}
\email{agata.checinska@fuw.edu.pl}
\affiliation{Institute of Theoretical Physics, University of Warsaw, Pasteura 5, 02-093 Warsaw, Poland}
\author{Alexander~R.~H.~Smith}
\email{a14smith@uwaterloo.ca}
\affiliation{Department of Physics \& Astronomy, University of Waterloo, Waterloo, Ontario Canada N2L 3G1}
\affiliation{Department of Physics \& Astronomy, Macquarie University, Sydney NSW 2109, Australia}
\author{Robert B. Mann}
\email{rbmann@uwaterloo.ca}
\affiliation{Department of Physics \& Astronomy, University of Waterloo, Waterloo, Ontario Canada N2L 3G1}
\affiliation{Perimeter Institute for Theoretical Physics, 31 Caroline Street North, Waterloo, Ontario N2L 2Y5, Canada}
\author{Andrzej Dragan}
\email{dragan@fuw.edu.pl}
\affiliation{Institute of Theoretical Physics, University of Warsaw, Pasteura 5, 02-093 Warsaw, Poland}
\date{\today}

\begin{abstract}
We study how an arbitrary Gaussian state of two localized wave packets, prepared in an inertial frame of reference, is described by a pair of uniformly accelerated observers. We explicitly compute the resulting state for arbitrarily chosen proper accelerations of the observers and independently tuned distance between them. To do so, we introduce a generalized Rindler frame of reference and analytically  derive the corresponding state transformation as a Gaussian channel. Our approach provides several new insights into the phenomenon of vacuum entanglement such as the highly non-trivial effect of spatial separation between the observers including sudden death of entanglement. We also calculate the fidelity of the two-mode channel for non-vacuum Gaussian states and obtain bounds on classical and quantum capacities of a single-mode channel. Our framework can be directly applied to any continuous variable quantum information protocol in which the effects of acceleration or gravity cannot be neglected.
\end{abstract}


\maketitle

\section{Introduction}

The vacuum state of a relativistic quantum field defined by an inertial (Minkowski) observer contains particles when observed from a uniformly accelerated frame of reference. As a consequence of this phenomenon, known as the Unruh effect \cite{Unruh1976}, the concept of particle, and in general any quantum state, becomes observer-dependent. Two major types of approaches have been developed in order to study the consequences of the Unruh effect. The first one involves implementing models of particle detectors that couple to the quantum field and can give different readouts depending on the detectors' motion \cite{BirrelDavies}. These detectors are typically described with various mutations of the Unruh-DeWitt model. Another approach involves a Bogolyubov transformation of a quantum state of the field on a given Cauchy surface from one basis of modes to another which corresponds to a transition between two reference frames \cite{Fuentes2005}. The advantage of the latter approach is that it is not constrained to a specific choice of the detection process and relies on a more general concept of a quantum state as the ultimate description of any physical system. However, very little progress has been made towards applying the Bogolyubov transformation method to quantum states other than the vacuum.

A seminal tool for investigating the transformation properties of non-vacuum field states was introduced in \cite{Bruschi2010} and involves the usage of so-called Unruh modes \cite{LD3}. Unfortunately this approach, followed by many authors, can easily lead one astray  \cite{papers}. Unruh modes are unphysical, delocalized and turn out to be coordinate-dependent objects that do not correspond to a uniformly accelerated observer moving with a specific proper acceleration. In particular, the results of \cite{papers} cannot be interpreted in terms of the dependence of the (entanglement of a) state on the acceleration of the involved observers \cite{Bruschi2010,LD1}.

Recently another method has been proposed to address the problem of the dependence of non-vacuum states on the motion of  observers \cite{LD1}. It has been already applied to a two-mode squeezed states of two localized wave packets of a scalar field \cite{LD2}. Another approach, based on calculating expectation values of localized field observables,  has also been proposed in order to study quantum communication between an inertial sender and an accelerated partner \cite{Downes2013}.

In this paper we present a general method for transforming any two-mode Gaussian state of two localized wave packets to an accelerated frame of reference corresponding to a pair of uniformly accelerated observers. In doing so we notice that this transformation can be represented as a Gaussian channel. Our approach, inspired by \cite{LD1}, not only allows us to give the explicit analytic expressions characterizing this channel with no approximations involved, but  it is also not constrained to the geometry of the Rindler chart. This enables us to study   scenarios in which the magnitude and direction of proper accelerations and the distance between the two non-inertial observers are not constrained in any way, i.e., they can be tuned independently. 

Our framework can readily be used for any quantum information protocol with bipartite continuous variable systems in the presence of strong accelerations. Moreover, due to the equivalence principle, our framework can be employed to describe the effect of gravity on Gaussian states. We discuss several applications of our scheme. First, we study how observations of the vacuum state are affected by the acceleration(s) of the observers, as well as by their spatial separation. Then, we investigate the transformation properties of two-mode squeezed thermal states. Finally, we discuss how single-mode Gaussian states transform to a uniformly accelerated reference frame and calculate bounds on classical and quantum capacities of the resulting channel.

Our paper is organized in the following way. In Sec.~\ref{theframework} we introduce our framework and show that the problem can be formulated in terms of a Gaussian channel. We completely characterize the resulting channel in Sec.~\ref{compchannel} for all possible magnitudes and directions of proper accelerations of the observers and their separations. In Sec.~\ref{ain} we show that the obtained results are coordinate-independent and discuss that fact in the context of the so-called entanglement degradation. In Sec.~\ref{modechoice} we discuss possible choices of the input and output modes of the channel. We apply our result to the input vacuum state in Sec.~\ref{vacent}, where we study how the extracted vacuum entanglement depends on the observers' accelerations  and their separation. In Sec.~\ref{secnonvacuum} we study the fidelity of the channel for a family of two-mode squeezed thermal states and in Sec.~\ref{singmode} we turn our attention to single-mode channels and compute bounds on their classical and quantum capacities. Finally, Sec.~\ref{concl} concludes our article. We present very detailed and pedagogical derivations of more technical aspects of our work in Appendices~\ref{computingnoise} and \ref{N12limits}.

\section{The framework\label{theframework}}

We consider a real scalar quantum field of a mass $m$ that satisfies the Klein-Gordon equation, $(\square+m^2)\hat{\Phi} = 0$, in $1+1$ dimensional Minkowski space-time (taking $c=\hslash=1$). This equation has a canonical scalar product associated with it that is preserved under  free evolution; in Minkowski coordinates it is defined as \cite{BirrelDavies}:
\begin{align}\label{SP}
(\phi_1,\phi_2) &= i\int_{\Sigma}\mbox{d}x \left(\phi_1^\star \partial_t\phi_2-\phi_2\partial_t\phi_1^\star \right),
\end{align}
where $\Sigma$ is a spacelike Cauchy surface  and the imaginary factor guarantees that $(\phi_1,\phi_2) = (\phi_2,\phi_1)^\star = -(\phi^\star_2,\phi^\star_1)$.
 
We consider two frames of reference: a Minkowski inertial observer and a uniformly accelerated Rindler observer with their corresponding complete decompositions of the field operator $\hat{\Phi}$ into  countable families of modes. The first mode decomposition involves wave packets $\phi_k$ with associated annihilation operators $\hat{f}_k$ that only contain positive frequencies with respect to the Minkowski timelike Killing vector field. The second decomposition involves an alternative family of wave packets $\psi_k$ with corresponding annihilation operators $\hat{d}_k$ that  only consist of positive frequencies with respect to the Rindler timelike Killing vector field. The two respective decompositions take the form:
\begin{equation}\label{phipsiexpa}
\hat{\Phi}=\sum_k\phi_k\hat{f}_k+\text{H.c.}=\sum_k\psi_k\hat{d}_k+\text{H.c.}
\end{equation}
We will investigate the scenario in which all the modes $\phi_k$ of the Minkowski frame decomposition are empty except the two modes\footnote{Although our framework can easily be generalized to an arbitrary number of modes.} $\phi_{\text{I}}$ and $\phi_{\text{II}}$ with their corresponding annihilation operators $\hat{f}_{\text{I}}$ and $\hat{f}_{\text{II}}$, which are in an arbitrary quantum state.  Our goal is to analyze the state of the quantum field with respect to the mode decomposition corresponding to the accelerated observers. However, we restrict our attention to two modes of the accelerated observer, $\psi_{\text{I}}$ and $\psi_{\text{II}}$, with $\hat{d}_{\text{I}}$ and $\hat{d}_{\text{II}}$ as their corresponding creation and annihilation operators\footnote{Note that at this stage the indices I and II do not necessarily indicate that the modes inhabit respective Rindler regions I and II.}. We discuss in detail how to specify these modes in Sec.~\ref{modechoice}. Although the remaining modes of the accelerated observer, $\psi_{k\neq\{\text{I},\text{II}\} }$, will not be empty, we choose to discard them.

We assume that the four modes introduced above, i.e., $\phi_{\text{I}}$, $\phi_{\text{II}}$, $\psi_{\text{I}}$, and $\psi_{\text{II}}$, are sufficiently localized in space in their respective frames in such a way that an approximate position can be attributed to each of them. As a consequence, a single proper acceleration can be approximately associated with each of the modes $\psi_{\text{I}}$ and $\psi_{\text{II}}$. Note that there exist certain constraints governing our ability to localize the wave packets. Since these  wave packets must be composed only of positive frequencies, their support in position space has to be non-compact.  We must also choose the envelope of the modes in such a way that the negative frequency contributions to their spectra can be safely ignored. We discuss these restrictions in detail in Sec.~\ref{modechoice}.

To investigate the scenario outlined above, we need to perform two quantum operations. The first one is a Bogolyubov transformation which corresponds to the change of mode basis from an inertial frame to an accelerated frame of reference. The second operation is ignoring all the modes in the accelerated frame except $\psi_{\text{I}}$ and $\psi_{\text{II}}$. A common feature of these two operations is that they transform Gaussian states into Gaussian states \cite{Holevo2001}. Therefore, without sacrificing any of the physically motivated questions, we limit our analysis to the Gaussian family of states which can be described in a very compact way\footnote{Gaussian states should not be confused with the spatial profiles of the modes they occupy, which can be arbitrary.} \cite{Gerardo2014}.

Gaussian states are completely described by first and second moments of their quadrature operators. We arrange these operators for the inertial modes as: 
\begin{align}\label{QO}
\hat{\vec{X}}^{(f)}=\left(\frac{\hat{f}_{\text{I}}+\hat{f}_{\text{I}}^{\dag}}{\sqrt{2}},\frac{\hat{f}_{\text{I}}-\hat{f}_{\text{I}}^{\dag}}{\sqrt{2}i},\frac{\hat{f}_{\text{II}}+\hat{f}_{\text{II}}^{\dag}}{\sqrt{2}},\frac{\hat{f}_{\text{II}}-\hat{f}_{\text{II}}^{\dag}}{\sqrt{2}i}\right)^T.
\end{align}
We define the vector of first moments, $\vec{X}^{(f)}$, and the matrix of second moments, $\sigma^{(f)}$, known as covariance matrix by: 
\begin{subequations}\label{CMFM}
\begin{align}
X^{(f)}_k & = \<\hat{X}_{k}^{(f)}\>,\label{FM}\\
\sigma^{(f)}_{kl} &=\left\langle \left\{\hat{X}^{(f)}_{k}- X_{k}^{(f)},\hat{X}^{(f)}_{l}- X_l^{(f)}\right\}\right\rangle\label{CM},
\end{align}
\end{subequations}
where the anti-commutator is defined as $\{\hat{A},\hat{B}\}=\hat{A}\hat{B}+\hat{B}\hat{A}$.
Similarly for the accelerated modes we define the vector of quadrature operators, first moments and the covariance matrix simply by replacing the letter $f$ by the letter $d$ in Eqs.~\eqref{QO} and \eqref{CMFM}.

The entire operation of transforming the state from one frame to another can be seen as the action of a noisy Gaussian channel \cite{Holevo2001}. This channel acts on the input Gaussian state of the modes, $\phi_{\text{I}}$ and $\phi_{\text{II}}$, defined in the inertial frame, and transforms them into the output Gaussian state of the modes, $\psi_{\text{I}}$ and $\psi_{\text{II}}$, defined in the accelerated frame. The transformation of the statistical moments of a general Gaussian state under the action of such a channel reads as \cite{Holevo2001}:
\begin{subequations}\label{GC}
\begin{align}
\vec{X}^{(d)} &=  M \vec{X}^{(f)},\label{GCFM}\\
\sigma^{(d)} & =  M\sigma^{(f)} M^T+N,
\label{GCCM}
\end{align}
\end{subequations}
where in our case $M$ and $N=N^T$ are $4\times 4$ real matrices and the matrix $N$ corresponds to the noise present in the quantum channel. Note that for a general Gaussian channel, the transformation of the first moments given by Eq.~\eqref{GCFM}, can include a constant displacement vector, which vanishes in our case. This is because a Bogolyubov transformation due to motion is always homogeneous. 

\begin{figure}
\centering
\includegraphics[width=1\linewidth]{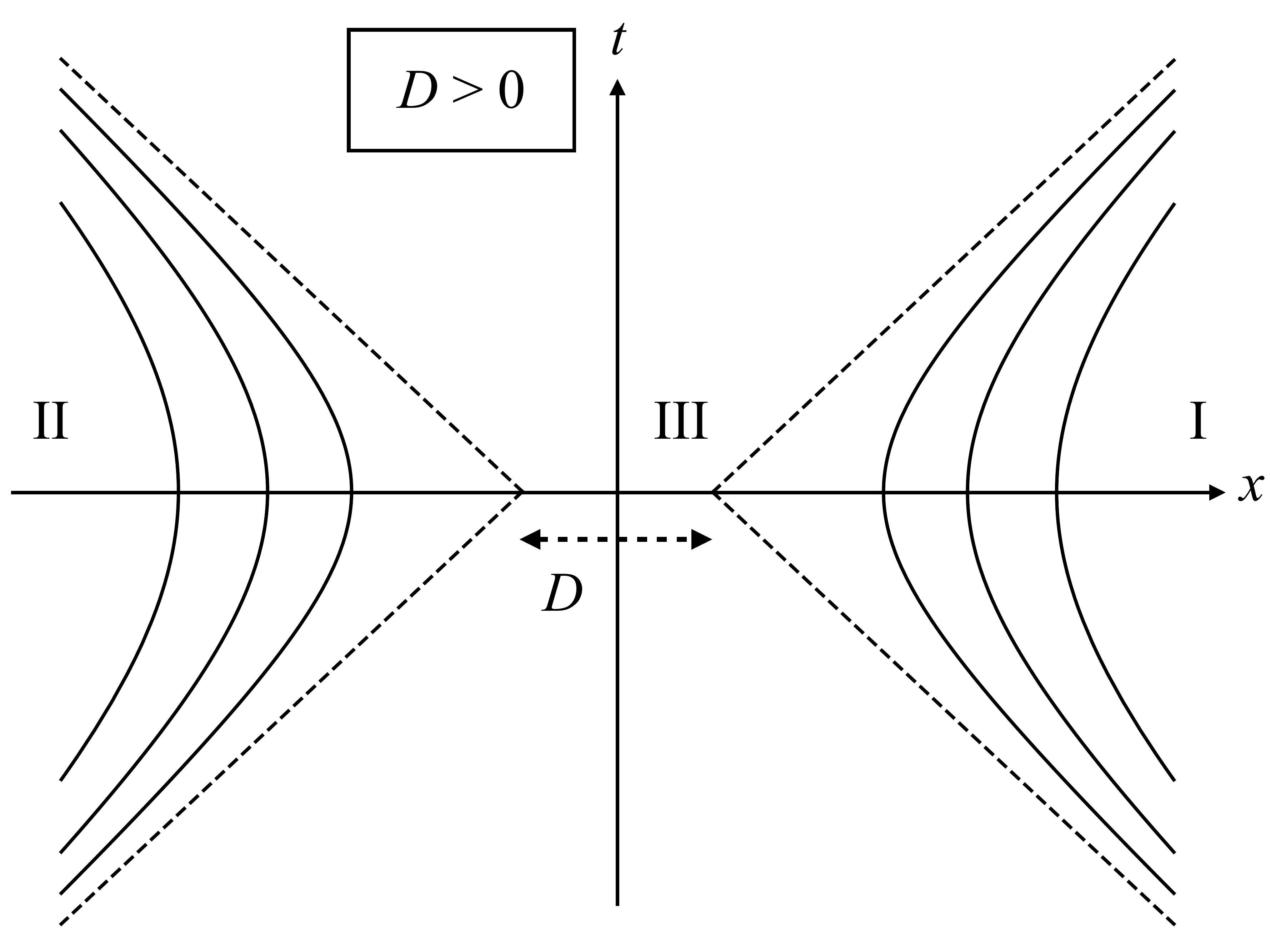}
\caption{When Rindler wedges I and II do not have a common apex and the two regions do not overlap.}\label{Dp}
\end{figure}

\begin{figure}
\centering
\includegraphics[width=1\linewidth]{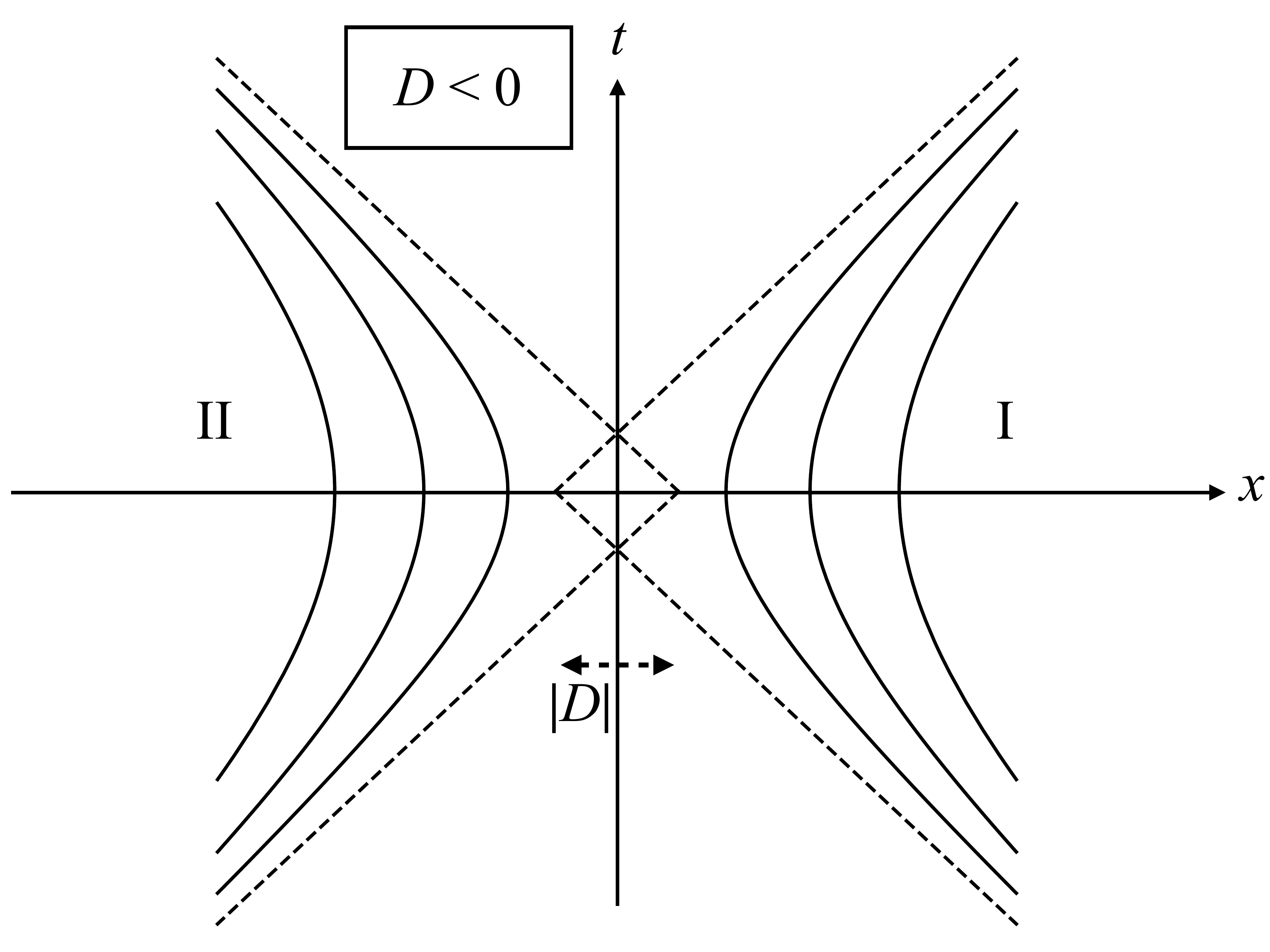}
\caption{When Rindler wedges I and II do not have a common apex and the two regions overlap.}\label{Dn}
\end{figure}

\begin{figure}
\centering
\includegraphics[width=1\linewidth]{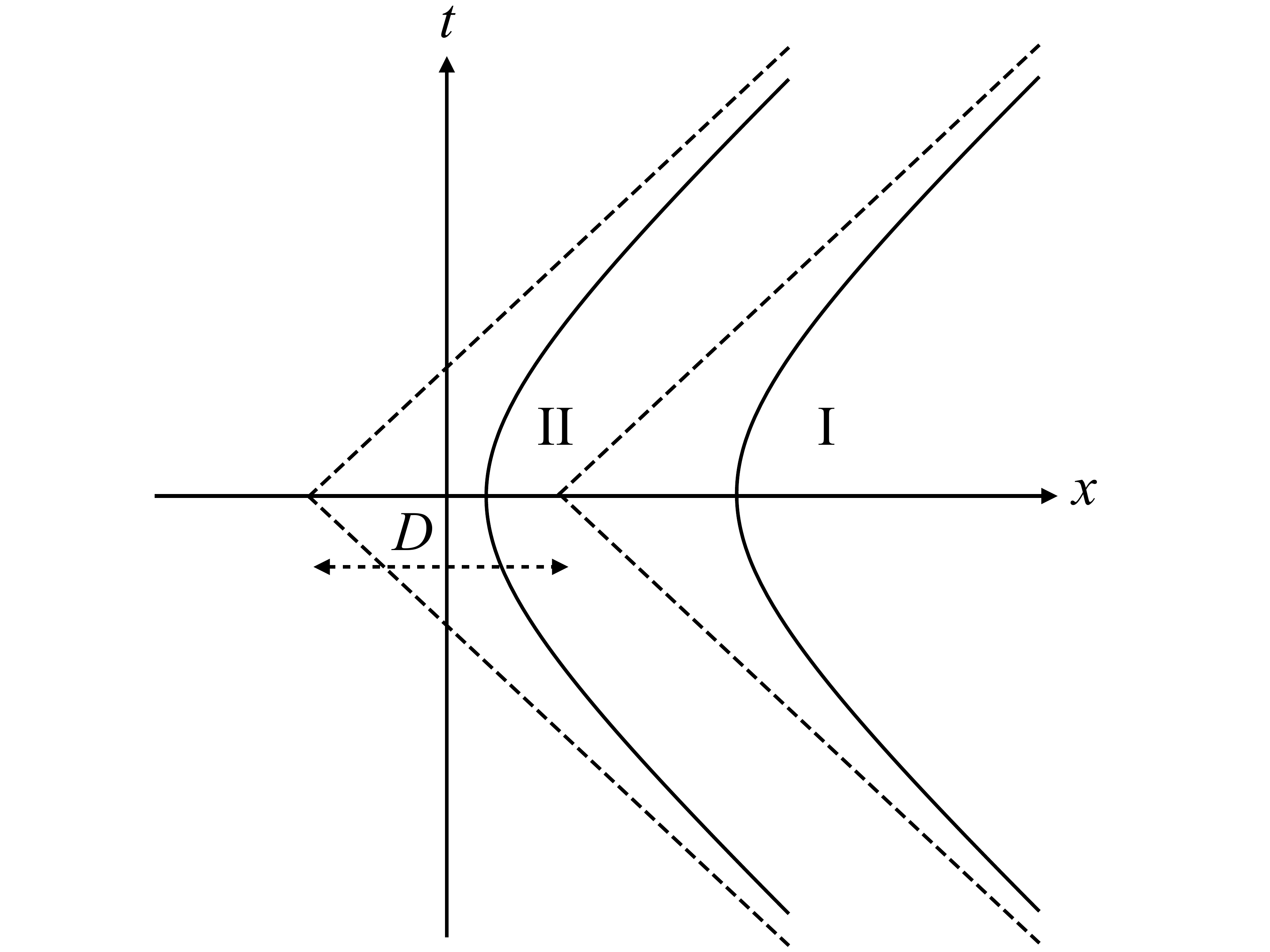}
\caption{Parallel accelerations with an additional distance $D$. In this figure we depict a situation wherein the two regions do not have a common apex and the two proper accelerations are in the same direction.}\label{Para}
\end{figure}

From Eqs.~\eqref{GC} it is  immediately clear that we only need to calculate the two matrices $M$ and $N$ to completely characterize the action of the quantum channel and consequently the effect of uniform acceleration on generic two-mode Gaussian states. Owing to this feature of Gaussian quantum channels our calculations are considerably simplified as compared to previous studies whilst being completely analytic and free of any approximations. Our approach enables us to study a general scenario wherein accelerated observers have arbitrary and well-defined positions and proper accelerations. Also we do not confine ourselves to the geometry of the Rindler chart, contrary to the common construction in which the two Rindler wedges meet at the origin. In doing so, we analyze the more general situation where the two accelerated observers and their corresponding Rindler wedges are separated by an arbitrary negative or positive distance as shown in Figs.~\ref{Dp} and \ref{Dn}, as well as the scenario in which both modes accelerate in the same direction, as depicted in Fig.~\ref{Para}.

\section{Computing the Gaussian quantum channel\label{compchannel}}

In this section we completely characterize the channel \eqref{GC} by calculating explicitly the matrices $M$ and $N$. In the literature, it is usually assumed that the distance between two counter-accelerated observers on the $t=0$ hypersurface is dependent on their proper accelerations, in which case their two Rindler wedges meet at the origin \cite{BirrelDavies}. In this paper we are interested in a more  general scenario in which the two Rindler wedges are separated by an additional distance $D$ and therefore the distance between the accelerated modes can be tuned independently of their proper accelerations \cite{Salton2015}. Note that the distance $D$ can be positive, as depicted in Fig.~\ref{Dp}, or negative as shown in Fig.~\ref{Dn}. In this section, first we analyze how the vector of first moments transforms according to Eq.~\eqref{GCFM} for an arbitrary distance $D$. This allows us to compute the matrix $M$ easily. Then, we compute the noise matrix $N$ for both cases of $D=0$, $D\neq 0$, as well as for the setting wherein the two observers are accelerating in the same direction.

Throughout the paper, we consider the transformation of the state of the field between the inertial and accelerated observers on the hypersurface $t=0$, for which the observers are relatively at rest. In a setup with a non-zero relative velocity one has to take into account an additional Doppler effect \cite{Downes2013}.

The mode functions considered in this work, $\phi_{\text{I}}$ and $\psi_{\text{I}}$ are assumed not to overlap with the mode functions $\phi_{\text{II}}$ and $\psi_{\text{II}}$:
\begin{equation}
\label{nooverlap}
(\phi_{\text{I}},\phi_{\text{II}}^{(\star)})=(\psi_{\text{I}}, \psi_{\text{II}}^{(\star)})=(\phi_{\text{I}}, \psi_{\text{II}}^{(\star)})=(\phi_{\text{II}},\psi_{\text{I}}^{(\star)})=0.
\end{equation}
Here, the symbol $^{(\star)}$ denotes the fact that the above equalities hold in the presence and absence of complex conjugation. As a consequence, for any $D$ we have: 
\begin{equation}
\left[\hat{f}_{\text{I}},\hat{f}_{\text{II}}^{(\dag)}\right] = \left[\hat{d}_{\text{I}},\hat{d}_{\text{II}}^{(\dag)}\right] = \left[\hat{f}_{\text{I}},\hat{d}_{\text{II}}^{(\dag)}\right] = \left[\hat{f}_{\text{II}},\hat{d}_{\text{I}}^{(\dag)}\right] =0.
\end{equation}
With the exception of the modes accelerated in the same direction as shown in Fig.~\ref{Para}, the index I will refer to region I of the Rindler chart and index II will refer to region II. Note that although we assume that the above modes do not overlap, the regions themselves may overlap (when $D<0$). 

Let us proceed with calculating the matrix $M$, which can be most easily studied by investigating the transformation properties of the first moments \eqref{GCFM}.    

\subsection{Computing matrix $M$} 

We begin by taking the scalar product $(\psi_\Lambda,\cdot)$, $\Lambda\in\{\text{I, II}\}$, of both decompositions of the quantum field operator as given in Eq.~\eqref{phipsiexpa} to obtain the Bogolyubov transformation relating the two sets of creation and annihilation operators. Using Eq.~\eqref{nooverlap} we obtain: 
\begin{equation}\label{dfbogo}
\hat{d}_\Lambda=\sum_k\left[(\psi_\Lambda,\phi_k)\hat{f}_k+(\psi_\Lambda,\phi_k^\star)\hat{f}_k^{\dag}\right].
\end{equation} 
Using the above equation together with the definitions \eqref{QO} and \eqref{FM}, we find the matrix $M$ which transforms the vector of first moments $\vec{X}^{(f)}$ from the inertial frame to the accelerated frame of reference according to \eqref{GCFM}. Since all the modes $\hat{f}_k$ are empty except for $k\in\{\text{I, II}\}$, for the purpose of calculating the first moments we can truncate the summation over $k$ in Eq.~\eqref{dfbogo} to two terms. After defining $
\ali=(\psii,\phii)$, 
$\bei=-(\psii,\phii^\star)$, 
$\alii=(\psiii,\phiii) $ and $\beii=-(\psiii,\phiii^\star)$, we obtain:
\begin{widetext}
\begin{equation}\label{M}
M=
\left(\begin{matrix}
\text{Re}(\ali-\bei) & -\text{Im}(\ali+\bei) & 0 & 0 \\
\text{Im}(\ali-\bei) & \text{Re}(\ali+\bei) & 0 & 0 \\
0 & 0 & \text{Re}(\alii-\beii) & -\text{Im}(\alii+\beii) \\
0 & 0 & \text{Im}(\alii-\beii) & \text{Re}(\alii+\beii) 
\end{matrix}\right).
\end{equation}
\end{widetext}
We emphasize that the derivation of the matrix $M$ is independent of the distance $D$ and it merely depends on the overlaps of the modes $\phi_{\text{I}}$, $\phi_{\text{II}}$ with the modes $\psi_{\text{I}}$, $\psi_{\text{II}}$. This is not the case for the noise matrix $N$ as we will see in the next section.

\subsection{Computing noise matrix $N$\label{compnoise}}

Before proceeding with the computation of the noise matrix $N$, let us introduce the natural coordinates for the accelerated frame of reference, namely the Rindler coordinates $\chi$ ($|\chi|$ is the proper distance from the event horizon) and $\eta$ with the two Rindler wedges separated by an extra distance $D$. The additional separation between regions can be both positive and negative, as shown in Figs.~\ref{Dp} and \ref{Dn}. When $D$ is positive, in addition to expanding the field in the Rindler wedges, one needs to introduce an additional region III, so that the field may be completely specified on a Cauchy surface. The case $D<0$ corresponds to the situation in which regions I and II of the Rindler chart partially overlap. This leads to the over-completeness of the basis spanned by the solutions of the field equation in the individual regions. The special case of $D=0$ corresponds to the standard definition of the Rindler coordinates $\chi$ and $\eta$. For an arbitrary separation $D$, we introduce a family of Rindler coordinates covering regions I and II parametrized with an arbitrary positive constant $a$ (interpreted as the proper acceleration of a trajectory for which $\eta$ is the proper time):
\begin{align}\label{rincoor}
t&=\chi\sinh(a\eta),\nonumber \\
x&=\chi\cosh(a\eta)\pm\frac{D}{2},
\end{align}
where $x$ and $t$ are Minkowski coordinates, the upper sign corresponds to the coordinates covering region I ($x>|t|+\frac{D}{2}$) for which $\chi>0$, and the lower sign corresponds to  region II ($x<-|t|-\frac{D}{2}$) for which $\chi<0$. We emphasize that the parameter $a$ should not be confused with the proper acceleration corresponding to the modes $\psi_{\text{I}}$ or $\psi_{\text{II}}$. The choice of $a$ is merely a convention and none of the physically relevant results should depend on it. In Sec.~\ref{ain} we show that indeed our results do not depend on the choice of this parameter.

The Klein-Gordon equation can be solved both in Minkowski coordinates and in Rindler coordinates \eqref{rincoor}. In the former, the normalized solutions are plane waves given by: 
\begin{equation}
\label{Minkowskimodes}
u_k = \frac{1}{\sqrt{4\pi\omega_k}} e^{i(kx-\omega_k t)},
\end{equation}
where $\omega_k = \sqrt{k^2+m^2}$, which satisfy the orthonormality relations $(u_k,u_l)=\delta{(k-l)}$, $(u_k^\star,u_l^\star)=-\delta(k-l)$ and $(u_k,u_l^\star)=0$. In Rindler coordinates the Klein-Gordon equation takes the following form \cite{Crispino2008, Dragan2011}:
\begin{equation}
\left[\frac{1}{a^2\chi^2}\frac{\partial^2}{\partial\eta^2}-\frac{\partial^2}{\partial\chi^2}-\frac{1}{\chi}\frac{\partial}{\partial\chi}+m^2\right]\hat{\Phi} = 0.
\end{equation}
Except for the special case $m=0$, corresponding to the massless field which will be discussed elsewhere~\cite{lorek}, the positive frequency solutions with respect to the timelike Killing vector fields in region I ($+\frac{\partial}{\partial \eta}$) and II ($-\frac{\partial}{\partial \eta}$) read as \cite{Crispino2008}:
\begin{align}\label{rinmodes}
w_{\text{I}\Omega}&= \sqrt{\frac{\sinh\left(\frac{\pi\Omega}{a}\right)}{\pi^2 a}} K_{i\frac{\Omega}{a}}\left( m\chi \right)e^{-i\Omega\eta}  \,\,\,\,\,\text{in I},\nonumber\\
w_{\text{II}\Omega}&= \sqrt{\frac{\sinh\left(\frac{\pi\Omega}{a}\right)}{\pi^2 a}} K_{i\frac{\Omega}{a}}\left( -m\chi \right)e^{i\Omega\eta} \,\,\,\,\,\,\,\,\text{in II},
\end{align}
and they vanish outside their respective regions. Here $K_{i\nu}(x)$ is the modified Bessel function of the second kind of pure imaginary order. Unlike the Minkowski solutions \eqref{Minkowskimodes} labeled by the wavevector $k$, the above modes are labeled by a strictly positive frequency index $\Omega>0$\footnote{This interesting discrepancy originates from the fact that only one of the two solutions of the Klein-Gordon equation in Rindler coordinates is (Dirac delta) normalizable, therefore the other one has to be discarded. A right-moving or left-moving mode solution on its own is unphysical because no freely evolving massive particle can escape to infinity when observed from a uniformly accelerated reference frame. In this frame a freely right-moving massive particle eventually turns back and becomes a left-mover. Similarly, any left-moving free particle must have been a right-mover in the past. Let us note that for the special case of $m=0$ no such restriction arises and both right-moving and left-moving (Dirac delta) normalizable solutions in the Rindler coordinates are allowed.\label{mfoot}}. The orthonormality relations for Rindler modes \eqref{rinmodes} read as: $(w_{\text{I}\Omega},w_{\text{I}\Xi})=\delta(\Omega-\Xi)$, $(w_{\text{I}\Omega}^\star,w_{\text{I}\Xi}^\star)=-\delta(\Omega-\Xi)$, and $(w_{\text{I}\Omega},w_{\text{I}\Xi}^\star)=0$. The same relations hold for the mode functions of region II, $w_{\text{II}\Omega}$. Furthermore, only for $D\geq 0$ we have $(w_{\text{I}\Omega},w_{\text{II}\Xi})=(w_{\text{I}\Omega}^\star,w_{\text{II}\Xi}^\star)=(w_{\text{I}\Omega},w_{\text{II}\Xi}^\star)=0$.

The field operator $\hat{\Phi}$ can always be decomposed in terms of the Minkowski modes $u_k$ and their corresponding annihilation operators $\hat{a}_k$ that satisfy canonical commutation relations $[\hat{a}_k,\hat{a}_l]=[\hat{a}_k\d,\hat{a}_l\d]=0$ and  $[\hat{a}_k,\hat{a}_l\d]=\delta(k-l)$. 

In the case $D=0$, the Rindler modes form an alternative complete basis for the field operator. However, this set of modes is no longer complete when $D>0$ and an additional component $\hat{\Phi}_{\text{III}}(D)$ of the field operator with support in region III must be introduced (Fig.~\ref{Dp}). Further, when $D<0$ the 
set of Rindler modes forms an overcomplete basis and in order to compensate for this, we also introduce an additional term $\hat{\Phi}_{\text{III}}(D)$: 
\begin{subequations}\label{minrindec}  
\begin{align}
\hat{\Phi} &=\int_{-\infty}^\infty \mathrm{d}k\, u_k \hat{a}_k+\text{H.c.}\label{mindec}\\
&=\int_0^\infty \mathrm{d}\Omega\,\left( \wi[\Omega]\bi[\Omega] + \wii[\Omega]\bii[\Omega] \right) +\text{H.c.} +\hat{\Phi}_{\text{III}}(D).
\label{rindec}
\end{align}
\end{subequations}
The explicit form of the term $\hat{\Phi}_{\text{III}}(D)$ is not relevant in our framework, except for the case when $D=0$ in which we have $\hat{\Phi}_{\text{III}}(0)=0$. In the decomposition \eqref{rindec}, $\bi[\Omega]$ and $\bii[\Omega]$ are the Rindler annihilation operators \cite{BirrelDavies}. The operators of region I satisfy the commutation relations $[\hat{b}_{\text{I}\Omega},\hat{b}_{\text{I}\Xi}]=[\hat{b}_{\text{I}\Omega}\d,\hat{b}_{\text{I}\Xi}\d]=0$ and $[\hat{b}_{\text{I}\Omega},\hat{b}_{\text{I}\Xi}\d]=\delta(\Omega-\Xi)$, and the same commutation relations hold for region II operators \cite{BirrelDavies}. Also when $D\geq 0$ the creation and annihilation operators of the two regions commute with each other, i.e., $[\hat{b}_{\text{I}\Omega},\hat{b}_{\text{II}\Xi}]=[\hat{b}_{\text{I}\Omega}\d,\hat{b}_{\text{II}\Xi}\d]=[\hat{b}_{\text{I}\Omega},\hat{b}_{\text{II}\Xi}\d]=0$, however, this is not true when $D<0$ due to the overcompleteness of the basis.

As mentioned earlier we assume that the wave packets $\psi_{\text{I}}$ and $\psi_{\text{II}}$ are only composed of positive frequencies with respect to the Rindler timelike Killing vector fields and therefore we have: 
\begin{subequations}\label{detrinbogo}
\begin{align}
\hat{d}_\text{I} &= \int_0^\infty \mathrm{d}\Omega\, (\psi_\text{I},\wi[\Omega])\bi[\Omega], \label{detrinbogoI}\\
\hat{d}_\text{II} &= \int_0^\infty \mathrm{d}\Omega\, (\psi_\text{II},\wii[\Omega])\bii[\Omega],\label{detrinbogoII}
\end{align}
\end{subequations}
where the normalization relations $\int \mathrm{d}\Omega\,|(\psi_\text{I},\wi[\Omega])|^2=1$  and $\int \mathrm{d}\Omega\,|(\psi_\text{II},\wii[\Omega])|^2=1$ hold. Although the individual Rindler mode operators $\bi[\Omega]$ and $\bii[\Xi]^{(\dagger)}$ do not commute when $D<0$, the wave packets' operators \eqref{detrinbogoI} and \eqref{detrinbogoII} are constructed in such a way that they commute for any $D$, i.e., $\left[\hat{d}_\text{I},\hat{d}^{(\dagger)}_\text{II}\right] = 0$. 
Our physical motivation for such a choice of the output wave packets $\psi_{\text{I}}$ and $\psi_{\text{II}}$ is to guarantee that the corresponding states are always orthogonal. We can always ensure this by sufficiently increasing the distance between the modes.
 
We now proceed with the investigation of the noise matrix $N$. As $N$ is independent of the input state of the channel, without any loss of generality, we choose this state to be the Minkowski vacuum, $|0\>_{\text{M}}$, with covariance matrix $\sigma^{(f)}_{\text{vac}}=\I$. Then using \eqref{GCCM} the noise matrix $N$ can be written as: 
\begin{equation}\label{N}
N = \sigma^{(d)}_{\text{vac}} - MM^T. 
\end{equation}
As we have already computed the matrix $M$ in Eq.~\eqref{M}, to compute the noise matrix $N$, we only need to compute the output state of the channel for the input Minkowski vacuum, i.e., $\sigma^{(d)}_{\text{vac}}$. In Appendix \ref{computingnoise}, we give a detailed derivation of that output state. After substituting the result into \eqref{N} we obtain:
\begin{align}
\label{noise}
N=&
\left(\begin{matrix}
1+N_{\text{I}}
&
0
&
\Re\,N_{\text{I,II}}^+
&
\Im\,N_{\text{I,II}}^-
\\
0
&
1+N_{\text{I}}
&
\Im\,N_{\text{I,II}}^+
&
-\Re\,N_{\text{I,II}}^-
\\
\Re\,N_{\text{I,II}}^+
&
\Im\,N_{\text{I,II}}^+
&
1+N_{\text{II}}
&
0
\\
\Im\,N_{\text{I,II}}^-
&
-\Re\,N_{\text{I,II}}^-
&
0
&
1+N_{\text{II}}
\end{matrix}\right)-M M^T,
\end{align}
where the diagonal terms are explicitly given as:
\begin{subequations}\label{Ns}
\begin{align}
N_{\text{I}} &=\int \mathrm{d}\Omega\, \frac{|(\psii,\wi[\Omega])|^2}{\sinh\left(\frac{\pi \Omega}{a}\right)}e^{-\frac{\pi \Omega}{a}},\label{N1}\\
N_{\text{II}} &=\int \mathrm{d}\Omega\, \frac{|(\psiii,\wii[\Omega])|^2}{\sinh\left(\frac{\pi \Omega}{a}\right)}e^{-\frac{\pi \Omega}{a}}.\label{N2}
\end{align}
\end{subequations}
These terms correspond to the well known Unruh noise, which is the average particle number that an observer would measure while accelerating through Minkowski vacuum \cite{LD1}. It is also worth pointing out that they are bounded from below by the $\beta_\Lambda$ coefficients characterizing the matrix $M$:
\begin{equation}
\label{nbeta}
2|\beta_\Lambda|^2 \leq  N_\Lambda
\end{equation}
for $\Lambda\in\{\text{I},\text{II}\}$, which is also proven in Appendix \ref{computingnoise}. The terms $N_{\text{I,II}}^\pm$ representing the existence of quantum correlations between regions I and II in the vacuum have a more complicated form:
\begin{widetext}
\begin{align}\label{N12forD}
N^\pm_{\text{I,II}}(D)=&\frac{1}{\pi a}\!\int\!\!\!\!\int\mathrm{d}\Omega\mathrm{d}\Xi\frac{(\psi_\text{I},\wi[\Omega])}{\sqrt{\sinh\left(\frac{\pi\Omega}{a}\right)\sinh\left(\frac{\pi\Xi}{a}\right)}}\times\nonumber\\
&\times\left[e^{\frac{\pi(\Omega-\Xi)}{2a}(1-\frac{D}{|D|})}(\psi_\text{II},\wii[\Xi])K_{i(\frac{\Omega-\Xi}{a})}(m|D|)
\pm
e^{\frac{\pi(\Omega+\Xi)}{2a}(1-\frac{D}{|D|})}(\psi_\text{II},\wii[\Xi])^\star K_{i(\frac{\Omega+\Xi}{a})}(m|D|)\right],
\end{align} 
\end{widetext}
and depend explicitly on the wedge separation $D$. For detailed calculation of Eqs.~\eqref{Ns}, \eqref{nbeta} and \eqref{N12forD}, we refer the readers to Appendix \ref{computingnoise}.

The above expressions describing the noise matrix $N$ together with Eq.~\eqref{M} completely characterize the Gaussian channel \eqref{GC} for an arbitrary choice of input and output modes as well as separation $D$. The Gaussian channel \eqref{GC} can now be readily used for any quantum information tasks involving continuous variable systems in accelerated motion or affected by gravity. It is worth pointing out that in the derivation of the above results we have used no approximations whatsoever. To complete the analysis, we study one more setting, namely the one in which the two output modes are accelerating in the same direction, as shown in Fig.~\ref{Para}. Before proceeding, let us briefly comment on the limiting cases of the result derived above. 

Let us study the asymptotic behavior of $N^\pm_{\text{I,II}}(D)$ given by Eq.~\eqref{N12forD} as $D\to 0$. For this purpose we employ the following property of the modified Bessel function: $\lim_{\epsilon\to0^+} K_{i\nu}(\epsilon)=\pi\delta(\nu)$, proven in Appendix \ref{N12limits}. Therefore in the limit of $|D|\to 0$ the first term of the integrand of Eq.~\eqref{N12forD} becomes proportional to $\delta\left({\frac{\Omega-\Xi}{a}}\right)$, while the second one vanishes (since the argument of delta is positive). This allows one to perform the integration over $\Xi$ in \eqref{N12forD} and leads to: 
\begin{align}
\lim_{|D|\to 0}N^\pm_{\text{I,II}}(D)&=\int\mathrm{d}\Omega\,\frac{(\psii,\wi[\Omega])(\psiii,\wii[\Omega])}{\sinh\left(\frac{\pi \Omega}{a}\right)}.\label{N12}
\end{align}
It can be seen that in order to witness the non-local correlations of the noise matrix $N$, not only do the two output modes have to be accelerated, but their spectra, $(\psi_\text{I},\wi[\Omega])$ and $(\psi_\text{II},\wii[\Omega])$, have to overlap. 

To show the asymptotic behavior of $N^\pm_{\text{I,II}}(D)$ as $D\to\infty$ we use the asymptotic form of the modified Bessel function for the large argument $|x|$: $K_{i\nu}(|x|)\approx \sqrt{\frac{\pi}{2|x|}}e^{-|x|}$ \cite{NIST}. Since in this limit the function becomes independent of the order $\nu$ and vanishes for large arguments, we can take the modified Bessel functions appearing in Eq.~\eqref{N12forD} outside the integral. Then in the limit of $|D|\to\infty $ the whole integral $N^\pm_{\text{I,II}}(D)$ vanishes:
\begin{align}
\lim_{|D|\to \infty}N^\pm_{\text{I,II}}(D)&=0.
\end{align}
We conclude that the non-local vacuum correlations must vanish at large distances.

Finally we turn our attention to the case wherein the trajectories corresponding to the two wave packets $\psi_\text{I}$ and $\psi_{\text{II}}$ are characterised by proper accelerations in the same direction, as shown Fig.~\ref{Para}. Note that although we label the two shifted Rindler wedges by I (right) and II (left), they both are the Rindler wedge I which is shifted by $\frac{D}{2}$ and $-\frac{D}{2}$ respectively. Let us emphasize that since we assume $\left[\hat{d}_\text{I},\hat{d}^{(\dagger)}_\text{II}\right] = 0$, we need to choose  $\psii$, $\psiii$ and $D$ in such a way that the two wave packets do not overlap. In Appendix \ref{computingnoise} we show that in this case only the off-diagonal elements of the noise matrix $N$ are changed, as compared with the expressions for the counter-accelerated case, and $N^\pm_{\text{I,II}}$ is replaced with:
\begin{widetext}
\begin{align}
\label{N12parallelfull}
N_{\text{I,II}}^{\!_{\,((\pm}}(D)&=\frac{1}{\pi a}\!\int\!\!\!\!\int\mathrm{d}\Omega\mathrm{d}\Xi\frac{(\psi_\text{I},\wi[\Omega])}{\sqrt{\sinh\left(\frac{\pi\Omega}{a}\right)\sinh\left(\frac{\pi\Xi}{a}\right)}}\times\nonumber\\
&\times\left[e^{\frac{\pi}{2a}[(\Omega-\Xi)-(\Omega+\Xi)\frac{D}{|D|}]}(\psi_\text{II},\wii[\Xi])K_{i(\frac{\Omega+\Xi}{a})}(m|D|)
\pm
e^{\frac{\pi}{2a}[(\Omega+\Xi)-(\Omega-\Xi)\frac{D}{|D|}]}(\psi_\text{II},\wii[\Xi])^\star K_{i(\frac{\Omega-\Xi}{a})}(m|D|)\right].
\end{align} 
\end{widetext}
Analogous to the previous scenario, we use the limit of the modified Bessel function $\lim_{\epsilon\to0^+} K_{i\nu}(\epsilon)=\pi\delta(\nu)$, from which we find:
\begin{align}
\lim_{|D|\to 0}N_{\text{I,II}}^{\!_{\,((\pm}}(D)&=\pm \int \mathrm{d}\Omega \frac{{(\psii,\wi[\Omega])(\psiii,\wii[\Omega])}^*}{\sinh(\frac{\pi\Omega}{a})}e^{\frac{\pi\Omega}{a}}.\label{N12parallel}
\end{align}
Using the asymptotic form of the modified Bessel function for large arguments, $K_{i\nu}(|x|)\approx \sqrt{\frac{\pi}{2|x|}}e^{-|x|}$, we find that when the separation $|D|$ between the wedges becomes large the noise correlations in the channel vanish:
\begin{align}
\lim_{|D|\to\infty}N_{\text{I,II}}^{\!_{\,((\pm}}(D)=0.
\end{align}
This last analysis completes the characterization of the Gaussian channel \eqref{GC} for all possible types of uniformly accelerated motion and arbitrary separations of the output modes. Let us note that Eqs.~\eqref{Ns}, \eqref{N12forD}, and \eqref{N12parallelfull} suggest that the noise matrix $N$ depends on the parameter $a$, which is of no physical relevance and its choice is merely a convention. In the next section, we will prove that  upon integration over $\Omega$ and $\Xi$ this dependence goes away and all these terms are in fact $a$-independent. 

\section{$a$-independence of the channel}\label{ain}
So far in our analysis we did not specify the input and output modes of the channel, $\phi_{\text{I}}$, $\phi_{\text{II}}$, $\psi_{\text{I}}$, and $\psi_{\text{II}}$. Let us now discuss how to choose these modes.

First of all, the four chosen modes have to contain only positive frequency contributions to their spectra, as defined in their respective reference frames. Mathematically, this means:
\begin{equation}
\label{nonegfreq}
(\phi_{\text{I}},u_k^\star) = (\phi_{\text{II}},u_k^\star) = (\psi_{\text{I}},w_{\text{I}\Omega}^\star) = (\psi_{\text{II}},w_{\text{II}\Omega}^\star) = 0
\end{equation}
for all $k$ and $\Omega$. The importance of this condition can be seen by considering the input state of the channel to be the Minkowski vacuum.  After taking the scalar product $(\phi_\Lambda,\cdot)$ with the quantities in
 Eqs. \eqref{phipsiexpa} and \eqref{mindec}, we get
\begin{align}
\hat{f}_{\Lambda}=\int\mathrm{d}k\,(\phi_{\Lambda},u_k)\hat{a}_k+(\phi_{\Lambda},u_k^{\star})\hat{a}_k^{\dag},
\end{align}
 from which we find that 
\begin{widetext}  
\begin{align}
{}_\text{M}\bra{0}\hat{f}_{\Lambda}^{\dag} \hat{f}_{\Lambda} \ket{0}{}_\text{M} 
&=\iint\mathrm{d}k \mathrm{d}l\,{}_\text{M}\bra{0}\left[(\phi_{\Lambda},u_k)^{\star}\hat{a}_k^{\dag}+(\phi_{\Lambda},u_k^{\star})^{\star}\hat{a}_k)\right] \left[(\phi_{\Lambda},u_l)\hat{a}_l+(\phi_{\Lambda},u_l^{\star})\hat{a}_l^{\dag})\right]\ket{0}{}_\text{M}\nonumber\\
&=\int\mathrm{d}k\,\left|(\phi_{\Lambda},u_k^{\star})\right|^2. 
\end{align}
\end{widetext}
 Therefore  abandoning the requirement~\eqref{nonegfreq} will result in particle production even for vanishing proper acceleration of the observer.

Second of all, we need the output states of the channel to be in principle measurable by a pair of uniformly accelerating detectors. It is crucial that individual points of any finite-size detector that undergoes a uniformly accelerated motion must experience different proper accelerations. One  can immediately see from Fig.~\ref{Dp} that the hyperbolae belonging to the Rindler chart are characterized by proper accelerations ranging from zero to infinity. To assign an approximate single proper acceleration ${\cal A}$ to an extended body, one must therefore ensure that its proper length $L$ is sufficiently small as compared to its proper distance from the event horizon, i.e., $L\ll \frac{1}{\cal A}$.

Furthermore, due to a finite size of any potential detector and limited duration of a possible measurement, the modes $\psii$ and $\psiii$ have to be sufficiently localized in space. We localize these modes to an extent that a single approximate proper acceleration ${\cal A}$ and proper distance from the event horizon $\frac{1}{\cal A}$ can be attributed to each of them. Consequently, these quantities can be operationally interpreted as those of a potential detector measuring the output modes. Such an interpretation is not possible when the considered modes are global, as in the commonly studied example of Unruh modes \cite{Bruschi2010, papers}.

Since no wave packet consisting of a purely positive frequency spectrum can be strictly localized to a finite region of space, we only consider modes that have infinite, but quickly vanishing tails. For instance, consider a Gaussian function that is approximately localized in space. With the right choice of its phase, this function  has tails long enough to ensure that the negative contribution to its spectrum is sufficiently small. Therefore the notion of localization that we use is weaker than that considered in the Reeh-Schlieder's theorem \cite{Reeh1961}. As a consequence, we do not face the interpretational problems that arise from that theorem.

In the following we show that for any choice of the modes according to the above requirements, the matrices $M$ and $N$ obtained previously are independent of the parameter $a$ appearing in the coordinate transformation \eqref{rincoor}. We first notice that the matrix $M$ given by Eq.~\eqref{M} merely consists of coordinate-independent scalar products between respective modes. Therefore, by the definition of a scalar product, $M$ cannot depend on the choice of $a$. In order to prove the same for the noise matrix $N$, let us define the most general form of the output modes. It is reasonable to only consider modes $\psi_{\Lambda \in \{{\rm I},{\rm II}\}}$ that are functions of coordinate-independent quantities, such as the proper distance from the event horizon, $|\chi|$, and the local proper time, $\tau = \frac{a}{{\cal A}_{\Lambda}}\eta$, i.e., the proper time measured within the region, where $\psi_{\Lambda}$ is localized. The mode can in principle also depend on its proper acceleration ${{\cal A}_{\Lambda}}$ via an arbitrary, but sufficiently localized function:
\begin{equation}\label{F1}
\psi_\Lambda = \psi_{\Lambda}(\chi,\tau, {{\cal A}_{\Lambda}}).
\end{equation}

To prove the $a$-independence of the noise matrix $N$ with the above choice, we first analyze the diagonal blocks of this matrix. To this end, we show that the dependence of  $N_{\Lambda}$ on $a$, as given by Eqs.~\eqref{Ns}, is merely apparent. Let us first calculate the scalar product $\left(\psi_{\Lambda},w_{\Lambda\Omega}\right)$ appearing in Eqs.~\eqref{Ns}. Using the definitions \eqref{SP} and \eqref{rinmodes} and changing variables from $(x,t)$ to $(\chi,\tau)$ we get: 
\begin{eqnarray}
\label{F2}
&&\left(\psi_{\Lambda},w_{\Lambda\Omega}\right)
=i\int_{\tau=0}\frac{\mathrm{d}\chi}{{\cal A}_{\Lambda}|\chi|}\left(\psi_{\Lambda}^\star\partial_{\tau}w_{\Lambda\Omega}-w_{\Lambda\Omega}\partial_{\tau}\psi_{\Lambda}^\star\right)\nonumber\\
&=&\sqrt{\frac{\sinh\left(\frac{\pi\Omega}{a}\right)}{\pi^2 a}}\int \frac{\mathrm{d}\chi}{{\cal A}_{\Lambda}|\chi|}\left(\pm{\cal A}_{\Lambda}\frac{\Omega}{a} \psi_{\Lambda}^\star - i\partial_\tau \psi_{\Lambda}^\star\right)K_{i\frac{\Omega}{a}}(m\chi)  \nonumber\\
&\equiv& \frac{1}{\sqrt{a}} F_{\Lambda}\left(\frac{\Omega}{a},{{\cal A}_{\Lambda}}\right),
\end{eqnarray}
where the upper sign corresponds to $\Lambda=\text{I}$, the lower sign to $\Lambda=\text{II}$ and  $F_{\Lambda}$ is a function specified by the choice of $\psi_{\Lambda}$. Here it is crucial that the function $F_{\Lambda}$ depends on $\Omega$ only through the ratio $\frac{\Omega}{a}$. We substitute this result into Eqs.~\eqref{N1} and \eqref{N2} to obtain: 
\begin{align}
N_{\Lambda}=\int \frac{\mathrm{d}\Omega}{a} \frac{\left|F_{\Lambda}(\frac{\Omega}{a},{{\cal A}_{\Lambda}})\right|^2}{\sinh\left(\frac{\pi \Omega}{a}\right)}e^{-\frac{\pi \Omega}{a}}.
\end{align}
Upon integration over the new variable $\frac{\Omega}{a}$, we conclude that $N_{\Lambda}$ is a function of proper acceleration ${{\cal A}_{\Lambda}}$ only and it is independent of the physically irrelevant parameter $a$.

In order to complete the proof of $a$-independence of the noise matrix $N$, we need to show that $N^\pm_{\text{I,II}}(D)$ given by Eq.~\eqref{N12forD} does not depend on $a$. Substituting the result of Eq.~\eqref{F2} into  Eq.~\eqref{N12forD} we find:
\begin{widetext}
\begin{align}
N^\pm_{\text{I,II}}(D)=&\frac{1}{\pi}\!\int\!\!\!\!\int\frac{\mathrm{d}\Omega}{a}\frac{\mathrm{d}\Xi}{a}\frac{F_{\text{I}}\left(\frac{\Omega}{a},{\cal A}_\text{I}\right)}{\sqrt{\sinh\left(\frac{\pi\Omega}{a}\right)\sinh\left(\frac{\pi\Xi}{a}\right)}}\times\nonumber\\
&\times\left[e^{\frac{\pi(\Omega-\Xi)}{2a}(1-\frac{D}{|D|})}F_{\text{II}}\left(\frac{\Xi}{a},{\cal A}_\text{II}\right)K_{i(\frac{\Omega-\Xi}{a})}(m|D|)
\pm
e^{\frac{\pi(\Omega+\Xi)}{2a}(1-\frac{D}{|D|})}F_{\text{II}}^\star\left(\frac{\Xi}{a},{\cal A}_\text{II}\right) K_{i(\frac{\Omega+\Xi}{a})}(m|D|)\right].
\end{align} 
\end{widetext}
Our claim that the above expression is independent of $a$ can be finally proved by changing the integration variables from $\Omega$ and $\Xi$ to $\frac{\Omega}{a}$ and $\frac{\Xi}{a}$. The proof proceeds analogously for the case of parallel accelerations.

We conclude that the properties of the channel \eqref{GC} are independent of parameter $a$ introduced in the Rindler transformation \eqref{rincoor} for any distance $D$ and any direction and magnitude of proper accelerations of the output modes. This means that for all purposes one can use the following coordinates $\chi$ and $\theta$, without introducing any parameter $a$, to characterize the uniformly accelerated observer in region I:
\begin{align}
t&=\chi \sinh \theta\nonumber\\ 
x&=\chi \cosh \theta.
\end{align}
Here $\chi$ is the proper distance from the event horizon and $\theta$ is a dimensionless temporal coordinate. 

This brings us to the question of how to interpret the so-called entanglement degradation commonly studied in the literature \cite{papers}. There it is argued that the entanglement of a state defined in the Minkowski frame is degraded due to the acceleration of the observer, and the amount of degradation depends on the parameter $a$, which we have shown to be physically irrelevant. Such effect is merely due to the fact that the input Unruh modes considered by these authors are not only global and unphysical, but also $a$-dependent themselves. The latter fact is commonly unnoticed, but careful inspection of the Unruh frequency parameterizing Unruh modes shows that this parameter implicitly depends on $a$ \cite{LD1,LD3}. We stress that the dependence of the entanglement degradation on the parameter $a$ studied in these works is merely due to the dependence of the initial state on $a$.

Another common misconception concerning entanglement degradation in accelerated frames is that this effect has something to do with the presence of Unruh particles that can obscure the entanglement of the input state. This Unruh noise, quantified in our approach by a contribution to the noise matrix $N$, will be shown to play a negligible role in the degradation process. The factor that is crucial for the degradation is the inevitable mode-mismatch between the input and output modes of the channel, characterized by the matrix $M$. The noise matrix $N$ plays a relevant role only for the studies of the vacuum entanglement and can be safely neglected when non-vacuum input states are considered. We show this in Sec.~\ref{vacent}.

\section{Choice of the modes\label{modechoice}}

In this section we discuss in detail possible choices of the input and output modes of the channel \eqref{GC}. Let us start with the modes $\phi_\Lambda$ with $\Lambda \in \{{\rm I},{\rm II}\}$ defined in the inertial reference frame that characterize the input state of the channel. We only need to define the mode functions and their first derivatives on the Cauchy surface $t = \eta = 0$. Similar to the choice of the modes in \cite{LD1,LD2,LD3}, one such choice that satisfies all our requirements discussed in the previous section is
\begin{align}
\label{inputgauss}
\phi_\Lambda(x,0)&= C e^{-2\left( \frac{x_0}{L}\log\frac{x}{x_0}\right)^2}\sin\left(\sqrt{\Omega_0^2-m^2} (x-x_0)\right), \nonumber \\
{\partial}_t \phi_\Lambda(x,0)&=-i  \Omega_0\phi_\Lambda(x,0),
\end{align}
where $x_0$ is the position around which the mode function is centered, $L$ is its width, and $C$ is a normalization constant. The frequency $\Omega_0$, about which the spectrum of the mode function is centered, has to be sufficiently large in order to effectively limit negative frequency contributions, i.e., $\Omega_0 \gg 1/L$. Additionally we impose an extra cut-off at zero frequency to completely eliminate the negative frequency contributions. Such a cut-off modifies the spatial profile given by \eqref{inputgauss} only a little. The parameters $L$ and $\Omega_0$ can be chosen independently for each region, i.e., $\Lambda\in \{\text{I}, \text{II}\}$. The exact form of the exponential envelope appearing in the definition \eqref{inputgauss} is chosen for later computational convenience, but to a good approximation it can be treated as a Gaussian: $e^{-2\left( \frac{x_0}{L}\log\frac{x}{x_0}\right)^2} \approx e^{-2\left(\frac{x-x_0}{L}\right)^2}$ as long as $|x-x_0| < L$. For the arguments $x$ that are further away from $x_0$ the tails of the chosen function vanish faster than the Gaussian tails.

\begin{figure}[t]
\includegraphics[width=1\linewidth]{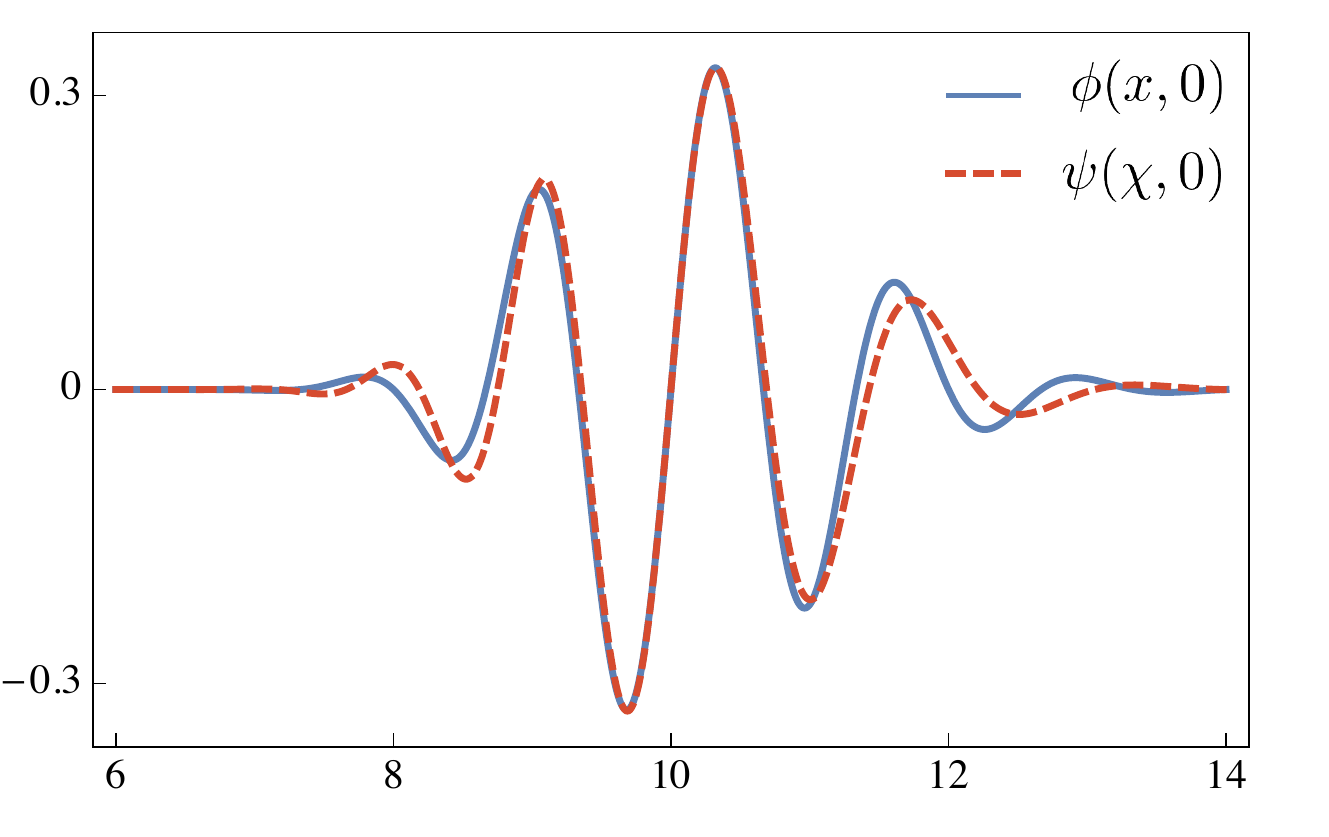}
\caption{\label{modescomparison}Comparison between the spatial modes $\phi$ and $\psi$ for the following choice of parameters: $x_0^{-1} = {\cal A}=0.1$, $L=2$, $\Omega_0\approx 5$, $m=0.1$.}
\end{figure}

The output modes $\psi_\Lambda$ are not determined by the choice of the input modes $\phi_\Lambda$, and their most general form is given by Eq.~\eqref{F1}. In this work we consider two natural  possible choices of the output modes. The first such choice corresponds to the situation where the accelerated modes $\psi_\Lambda$ change with respect to the inertial modes $\phi_\Lambda$ in analogy with the behavior of the eigenmodes of a uniformly accelerated cavity. Consider a field in a resting inertial cavity of length $L$, satisfying Dirichlet boundary conditions. The eigenmodes  are standing waves in the Minkowski coordinates $x$ and $t$ that vanish outside the cavity. Our choice of the input modes \eqref{inputgauss} corresponds to the modes of a cavity with a sharply vanishing characteristic function replaced with a smooth exponential envelope quickly vanishing outside the cavity. Suppose that the cavity uniformly accelerates while its proper length $L$ remains unchanged, which means that the left mirror of the cavity accelerates differently from the right one. The eigenmodes of the accelerated cavity are combinations of the modified Bessel functions in Rindler coordinates $\chi$ and $\eta$. Therefore a natural choice of the output modes $\psi_\Lambda$ is similar to \eqref{inputgauss} but with the position $x$ replaced by the proper distance $\chi$, $t$ replaced by proper time at the center of the mode, $\tau$, and the trigonometric function replaced by the following combination of modified Bessel functions of the first kind, $I_{i\nu}(x)$ \cite{Dragan2011}:
\begin{equation}
\label{spatialmode}
f(\chi)= \Im \left[I_{-i\frac{\Omega_0}{\cal A}}(m|x_0|)I_{i\frac{\Omega_0}{\cal A}}(m|\chi|)\right].
\end{equation}
Therefore at Cauchy surface $\eta=\tau=0$ we have:
\begin{align}
\label{outputgauss}
&\psi_\Lambda(\chi,0)=C'  e^{-2\left( \frac{x_0}{L}\log\frac{\chi}{x_0}\right)^2}f(\chi), \nonumber \\
&{\partial}_\tau \psi_\Lambda(\chi,0)=\mp i  \Omega_0\psi_\Lambda(\chi,0),
\end{align}
where $C'$ is the normalisation constant, and the upper(lower) sign refers to $\Lambda=\text{I}(\text{II})$. Here, we also impose a zero-frequency cut-off to eliminate any remnants of the negative frequency contributions. We can additionally choose  $|x_0| = \frac{1}{\cal A}$, so that the input and output modes are localized in the same region of space. We will refer to these modes as being passive output modes, since the dependence of the modes \eqref{outputgauss} on proper acceleration is analogous to the way cavity eigenmodes are affected by the proper acceleration of the cavity. We show how the acceleration affects the output modes in Fig.~\ref{modescomparison}, where we compare the modes given by Eqs.~\eqref{inputgauss} and \eqref{outputgauss}. The parameters are chosen such that both modes are sufficiently localized, $\phi$ contains only positive frequency spectrum in the Minkowski frame, and $\psi$ contains only positive frequency spectrum in the Rindler frame.

The second natural choice of the output modes $\psi_\Lambda$ involves choosing these modes for each proper acceleration ${\cal A}_\Lambda$ separately. For example, the accelerated observer could optimize the fidelity of the considered channel over all possible choices of the output modes for each proper acceleration separately. This can be done by selecting the modes $\psi_\Lambda$ in such a way that the overlaps $\alpha_\Lambda=\left(\psi_\Lambda,\phi_\Lambda\right)$ in the matrix $M$, given by Eq.~\eqref{M}, are maximized. As was proved in \cite{LD2}, the optimum choice for this purpose is given by:
\begin{align}
\label{activemodes}
\psi_\Lambda=\frac{\int\mathrm{d}\Omega \left(w_{\Lambda \Omega},\phi_\Lambda\right) w_{\Lambda \Omega}}{\sqrt{\int\mathrm{d}\Omega |\left(\phi_\Lambda,w_{\Lambda \Omega}\right)|^2}}.
\end{align}
These modes $\psi_\Lambda$ can be simply understood as the modes $\phi_\Lambda$ with the spectra truncated to strictly positive frequency Rindler component. For the above choice, the maximized $\alpha_\Lambda$ coefficients are given as: 
\begin{equation}
\label{alphaactive}
\alpha_\Lambda=\sqrt{\int\mathrm{d}\Omega |\left(\phi_\Lambda,w_{\Lambda \Omega}\right)|^2},
\end{equation}
and $\beta_\Lambda$ coefficients turn out to be: 
\begin{align}
\beta_{\Lambda}=\alpha_\Lambda^{-1}\int\mathrm{d}\Omega \left(\phi_\Lambda,w_{\Lambda \Omega}\right)(\phi_\Lambda,w_{\Lambda \Omega}^\star).
\end{align}
We will refer to the modes chosen above as active output modes. As we will discuss in Sec.~\ref{secnonvacuum}, these active modes characterize a channel of a maximum possible fidelity.

\section{Entanglement of the vacuum\label{vacent}}

In this section, we analyze the effect of the noisy channel \eqref{GC} on the Minkowski vacuum as an input state. Therefore we have $\vec{X}^{(f)}=\vec{0}$ and $\sigma^{(f)} = \openone$, for which Eq.~\eqref{GC} reduces to:
\begin{align}
\label{vacuumout}
\vec{X}^{(d)} &=  0,\nonumber \\
\sigma^{(d)} & =
\left(\begin{matrix}
1+N_{\text{I}}
&
0
&
\Re \,N_{\text{I,II}}^+
&
\Im\,N_{\text{I,II}}^-
\\
0
&
1+N_{\text{I}}
&
\Im\,N_{\text{I,II}}^+
&
-\Re\,N_{\text{I,II}}^-
\\
\Re\,N_{\text{I,II}}^+
&
\Im\,N_{\text{I,II}}^+
&
1+N_{\text{II}}
&
0
\\
\Im\,N_{\text{I,II}}^-
&
-\Re\,N_{\text{I,II}}^-
&
0
&
1+N_{\text{II}}
\end{matrix}\right),
\end{align}
where $N_{\text{I}}$ and $N_{\text{II}}$ are given by Eqs.~\eqref{N1} and \eqref{N2}. The off-diagonal terms $N^\pm_{\text{I,II}}$ are given in the counter-accelerating case by Eq.~\eqref{N12forD} and in the co-accelerating case by Eq.~\eqref{N12parallelfull}. Let us note that the results presented in this section also hold for coherent states at the input, for which $\vec{X}^{(f)} \neq  0$. Here, we focus on the entanglement present in the output state of the channel. To do so, we quantify the amount of entanglement using the logarithmic negativity ${\cal E_N}$. This quantity is a measure of distillable entanglement and is particularly easy to compute for any two-mode Gaussian state \cite{Gerardo2004}. For the output state $\sigma^{(d)}$ given by Eq.~\eqref{vacuumout} the logarithmic negativity is equal to: 
\begin{align} 
{\cal E_N} = \max\left\{0,-\log\sqrt{\frac{\Delta-\sqrt{\Delta^2-4\det \sigma^{(d)}}}{2}}\right\},
\end{align}
where $\Delta\equiv (1+N_{\text{I}})^2 +  (1+N_{\text{II}})^2 + 
\Re\,N_{\text{I,II}}^+ \Re\,N_{\text{I,II}}^- +
\Im\,N_{\text{I,II}}^+ \Im\,N_{\text{I,II}}^-$. 

In order to calculate ${\cal E_N}$ we need to explicitly compute $N_{\text{I}}$, $N_{\text{II}}$ given by Eqs.~\eqref{N1}, \eqref{N2}, and $N^\pm_{\text{I,II}}$ given by Eq.~\eqref{N12forD} for the counter-accelerating case or Eq.~\eqref{N12parallelfull} for the co-accelerating case. These expressions involve multiple integrations, which we evaluate numerically. The integrals \eqref{N1} and \eqref{N2} do not depend on $D$ and are relatively easy to compute. The most challenging term to compute is $N^\pm_{\text{I,II}}$ that is responsible for the non-local correlations of the vacuum state. This term involves quadruple integrations: two integrations over spatial coordinate given by the ovelaps $(\psi_\text{I},\wi[\Omega])$ and $(\psi_\text{II},\wii[\Xi])$ and the remaining double integral over Rindler frequencies $\Omega$ and $\Xi$. The last two integrations are particularly hard to evaluate numerically due to a rapidly oscillating integrand. 

\begin{figure}[t]
\includegraphics[width=1\linewidth]{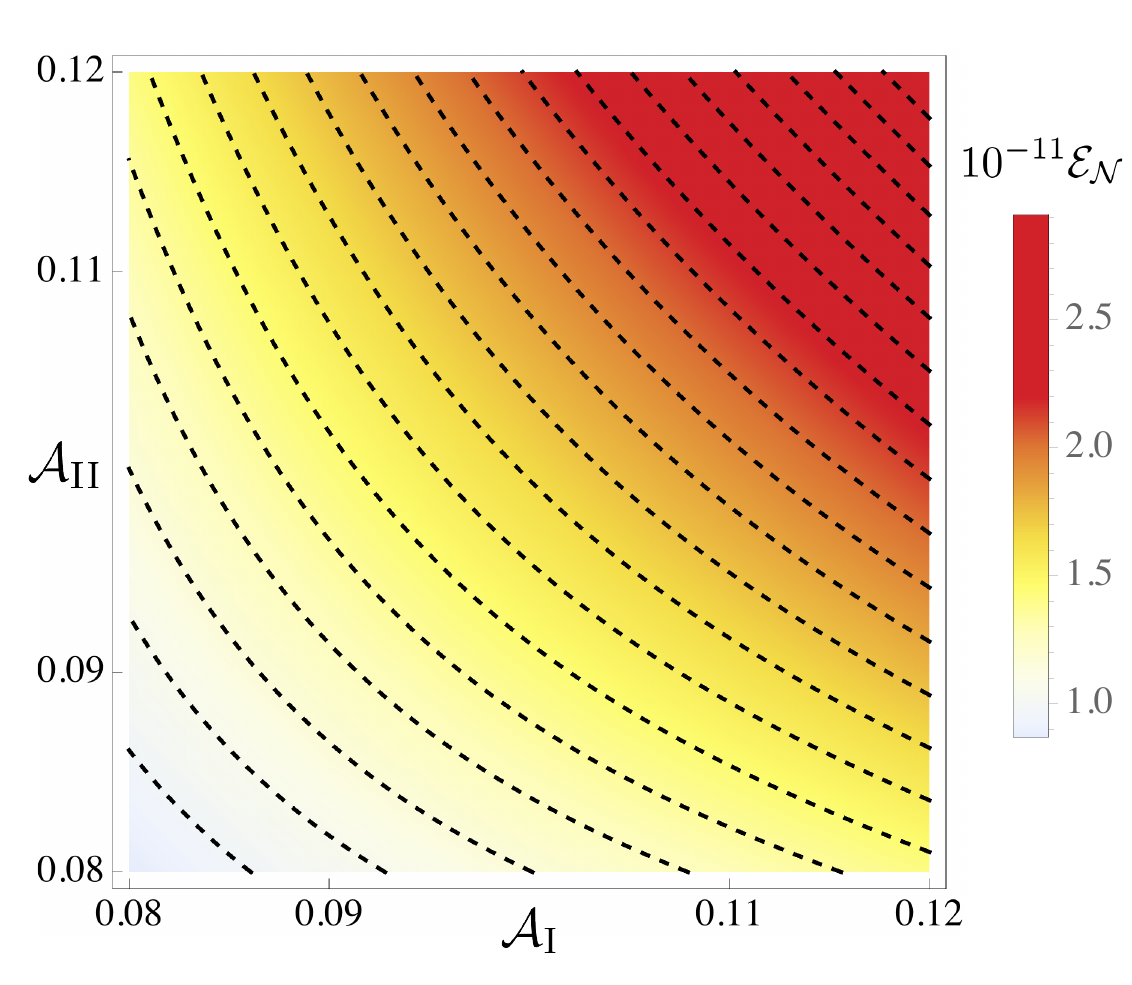}
\caption{\label{D0_vs_Accelerations}Logarithmic negativity of the Minkowski vacuum for two counter-accelerated modes, as a function of their proper accelerations for $D=0$. We have chosen $L=2$, $m=0.1$ and $\Omega_0\approx 5$. }
\end{figure}

\begin{figure}[t]
\includegraphics[width=1\linewidth]{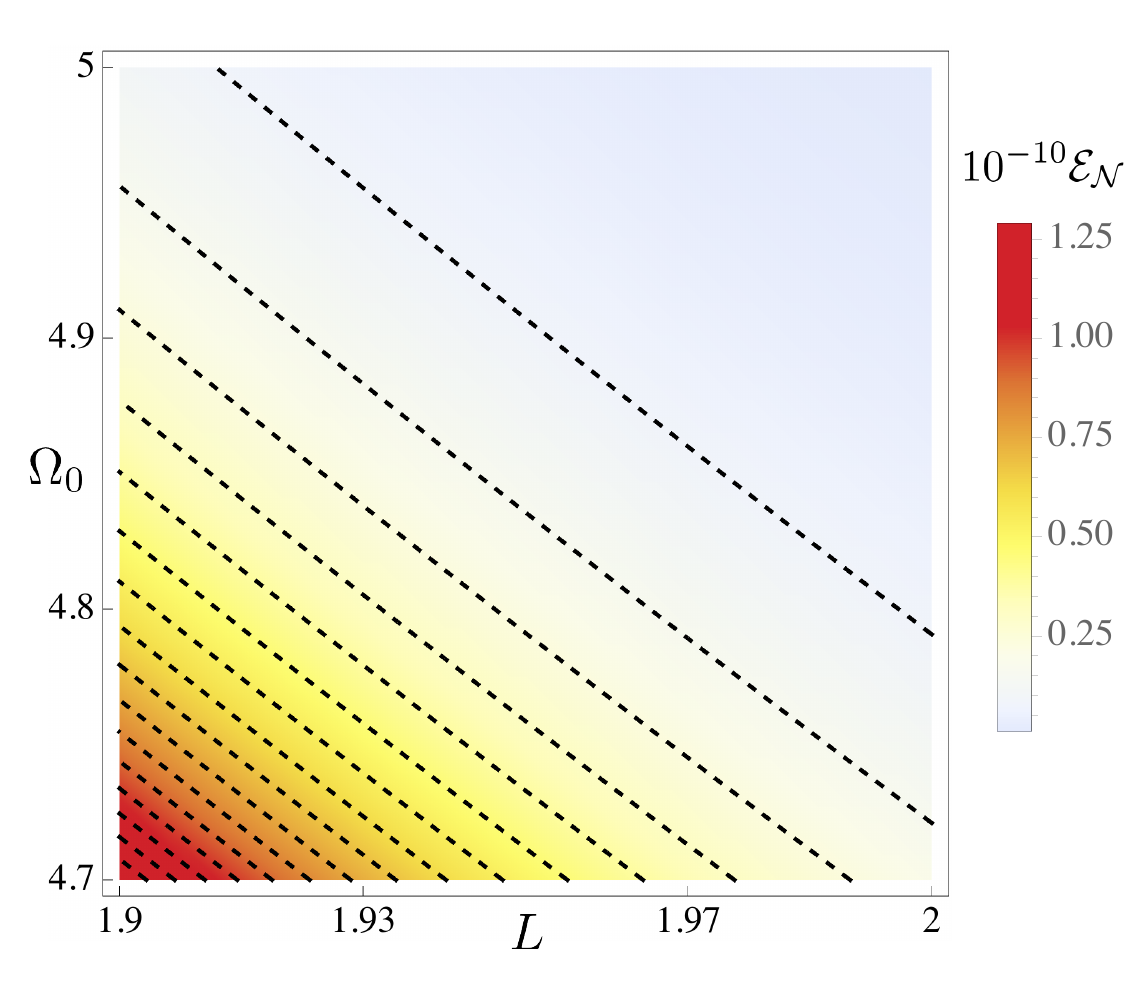}
\caption{\label{Negativity_vs_LOmega}Logarithmic negativity of the Minkowski vacuum for two identical counter-accelerated modes characterized by the same proper acceleration ${\cal A}=0.1$ as a function of $L$ and $\Omega_0$. We have also fixed $D=0$ and $m=0.1$.}
\end{figure}

\begin{figure}[t]
\includegraphics[width=1\linewidth]{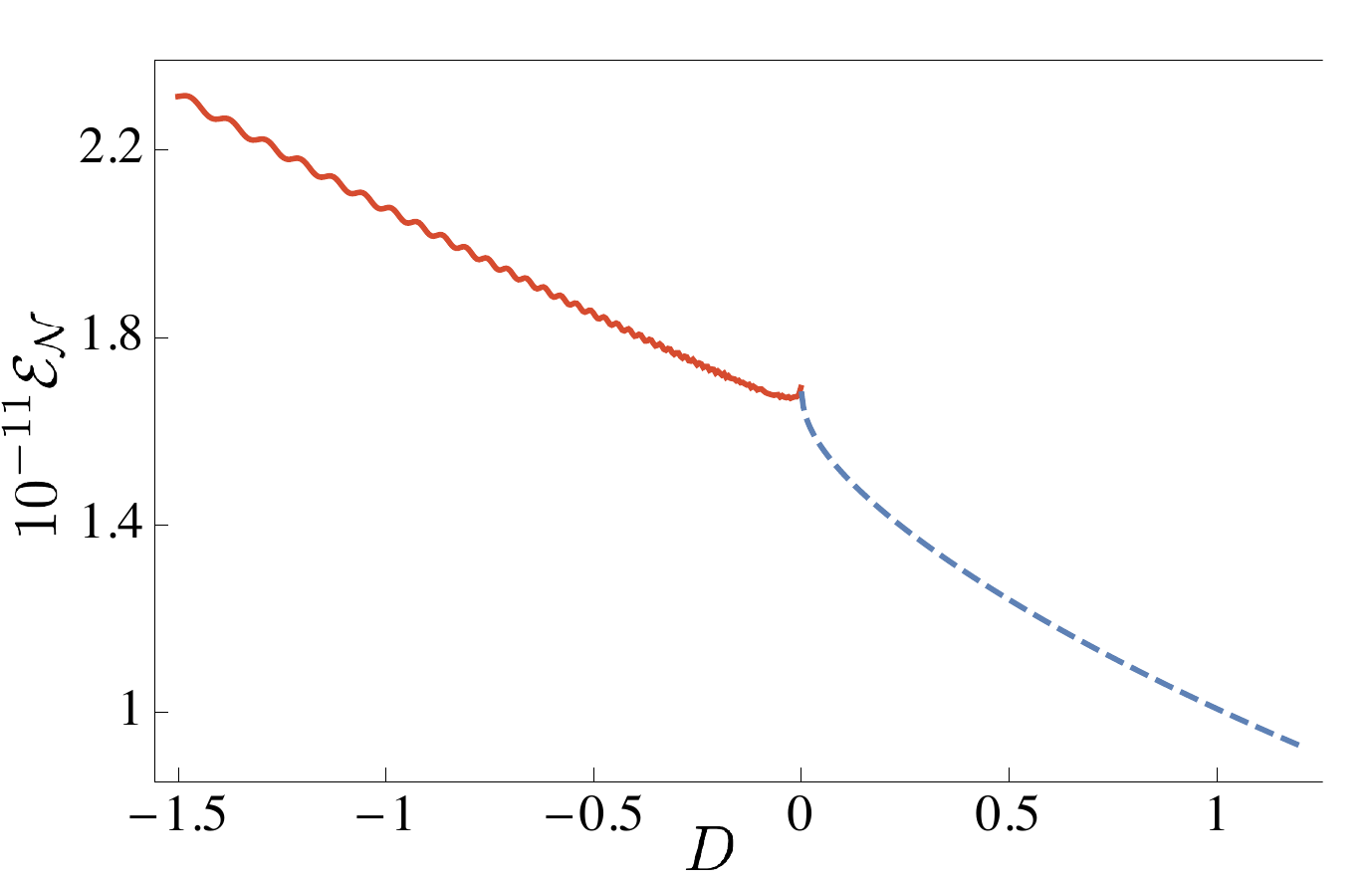}
\includegraphics[width=1\linewidth]{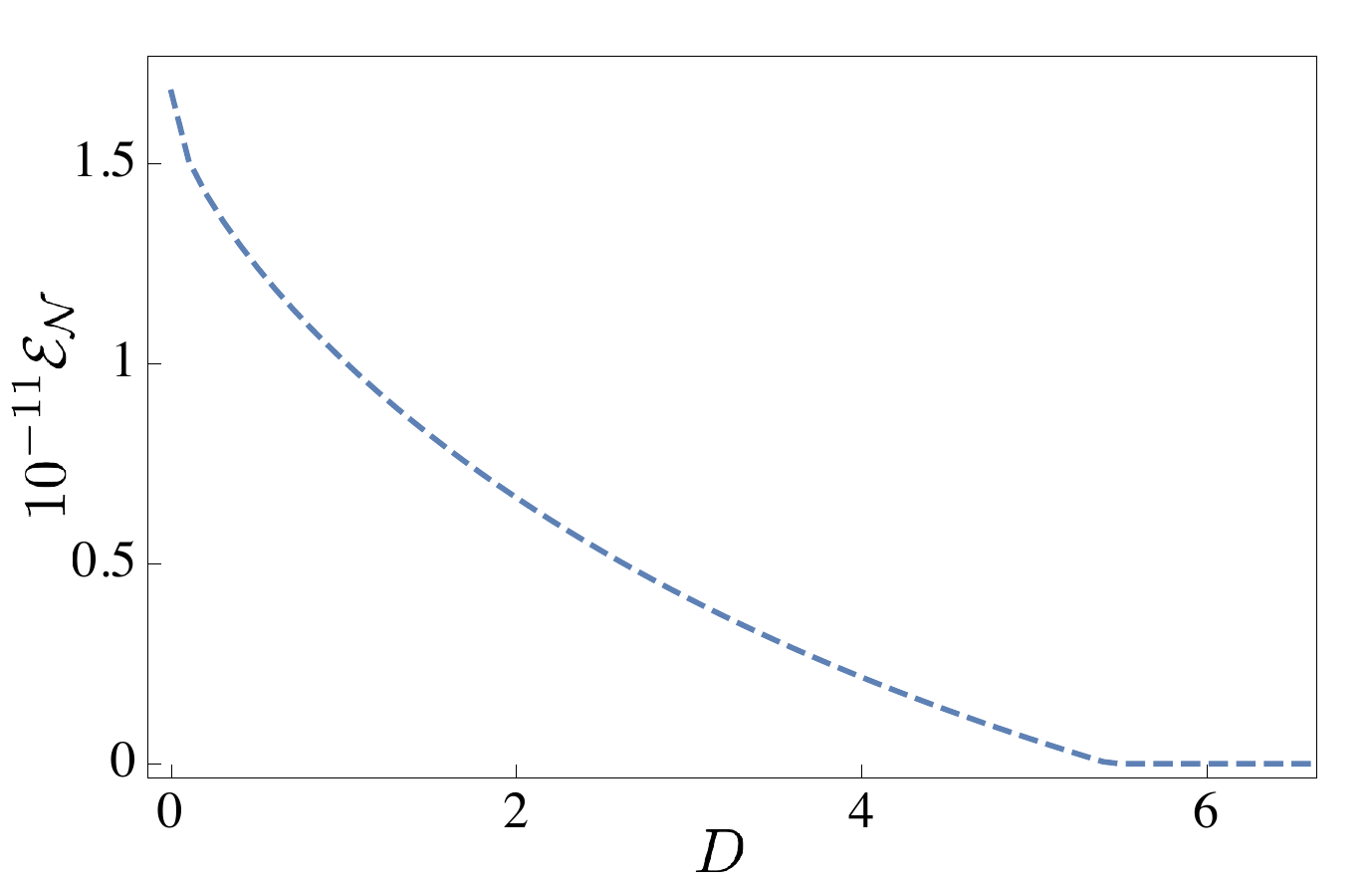}
\caption{\label{Negativity_vs_D}Logarithmic negativity of the Minkowski vacuum for two counter-accelerated modes, for fixed and equal proper accelerations ${\cal A}_{\text{I}}={\cal A}_{\text{II}} = 0.1$ as a function of the distance $D$. We have chosen $L=2$, $m=0.1$, and $\Omega_0\approx 5$. In the upper plot we focus on small wedge separations $D\approx 0$ and the lower plot shows the behavior of the negativity in a larger scale $D>0$. Solid lines correspond to $D<0$ and the dashed lines correspond to $D>0$.}
\end{figure}

\begin{figure}[t]
\includegraphics[width=1\linewidth]{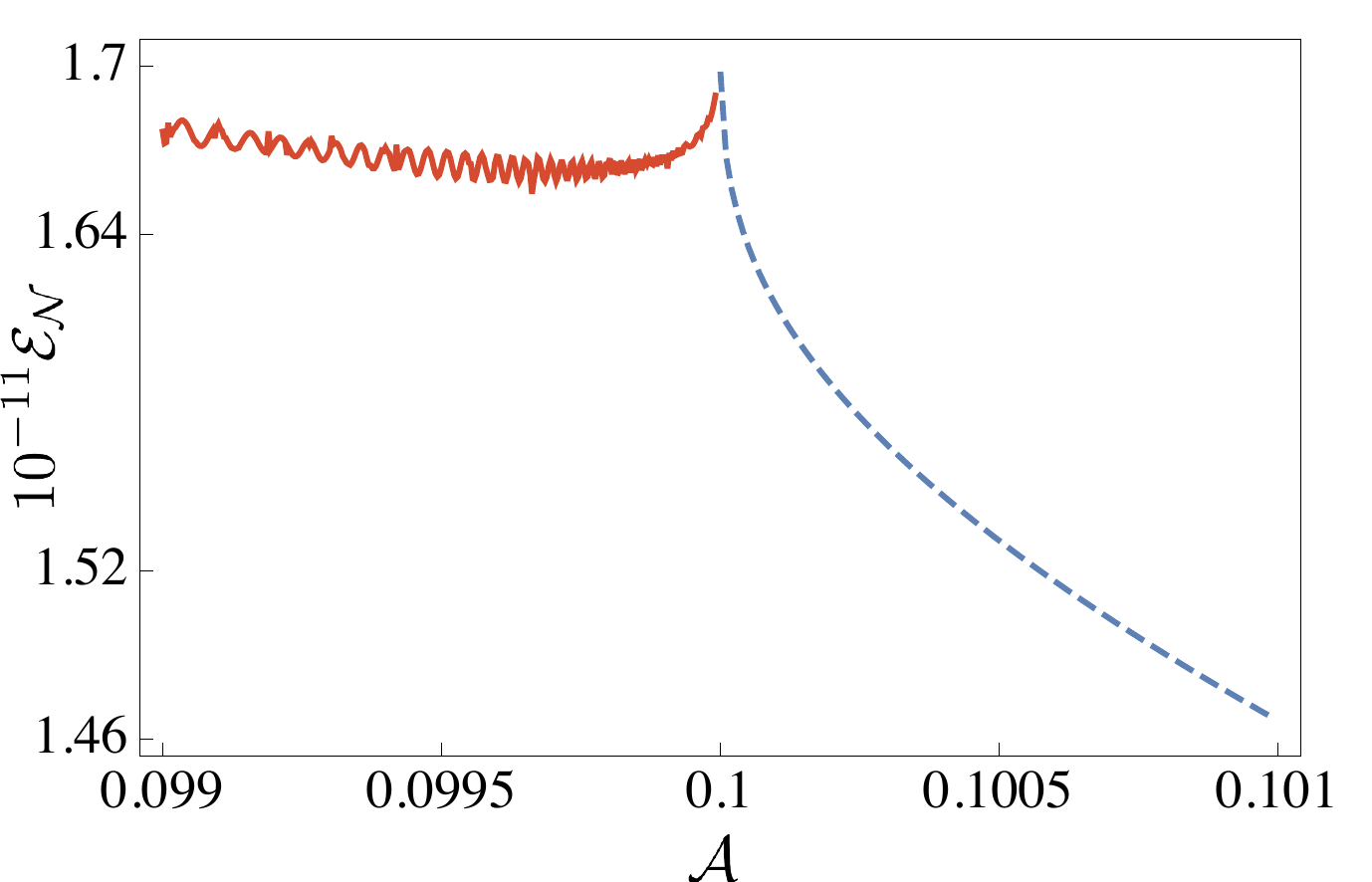}
\includegraphics[width=1\linewidth]{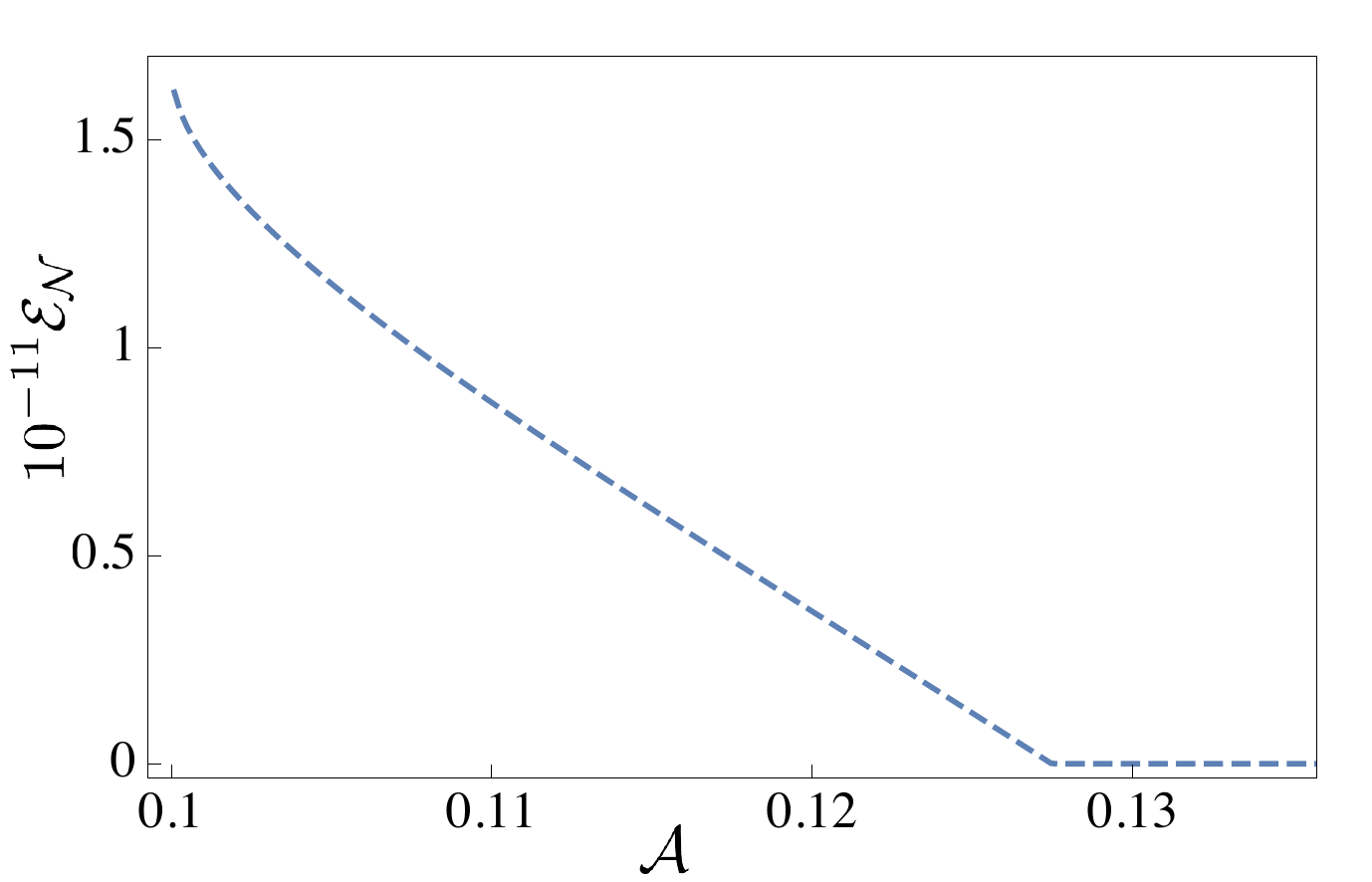}
\caption{\label{Negativity_fixed_distance}Logarithmic negativity of the Minkowski vacuum for two counter-accelerated modes, as a function of proper acceleration ${\cal A}_{\text{I}} = {\cal A}_{\text{II}}\equiv {\cal A}$ for $D = 20-\frac{2}{{\cal A}}$ such that the separation between modes is fixed and equal to $20$. We have chosen $L=2$, $m=0.1$, and $\Omega_0\approx 5$. In the upper plot we focus on small wedge separations $D\approx 0$ and the lower plot shows the behavior of the negativity in a larger scale, when $D>0$. Solid lines correspond to $D<0$ and the dashed lines correspond to $D>0$.}
\end{figure}

Note that the computations in the counter-accelerating case are straightforward for $D=0$, since Eq.~\eqref{N12forD} reduces to a simple integral \eqref{N12}. The results are presented in a series of Figures: \ref{D0_vs_Accelerations}-\ref{Negativity_fixed_distance}.

In Fig.~\ref{D0_vs_Accelerations}, we compute the logarithmic negativity for the state of the output modes characterized by the same width $L=0.1$ and central frequency $\Omega_0\approx 5$ as a function of the proper accelerations ${\cal A}_{\text{I}}$ and ${\cal A}_{\text{II}}$. Here, we show the results in the case of $D=0$. As discussed in section \ref{ain}, we only consider proper accelerations such that ${\cal A}L\ll 1$. Fig.~\ref{D0_vs_Accelerations} shows that the entanglement present in the vacuum state, as seen by the accelerating modes $\psi_\Lambda$ is an increasing function of both proper accelerations ${\cal A}_{\text{I}}$ and ${\cal A}_{\text{II}}$, which is consistent with the results known in the literature \cite{LD1}. Next, in Fig.~\ref{Negativity_vs_LOmega} we investigate how the entanglement of the output state is affected by the central frequency of the modes, $\Omega_0$, and their width, $L$. In Fig.~\ref{Negativity_vs_LOmega} we consider both modes to be identical and characterized by a fixed proper acceleration ${\cal A}_{\text{I}} = {\cal A}_{\text{II}} = 0.1$, and the separation $D=0$. It turns out that the extracted entanglement is larger in the infrared end of the spectrum and it also increases with decreasing $L$. So far our results are not very surprising and find confirmation in the literature on entanglement of the vacuum \cite{LD1}.

A more challenging and surprising result is shown in Fig.~\ref{Negativity_vs_D}, where we compute the logarithmic negativity for a pair of identical, counter accelerating output modes characterized by the same proper accelerations ${\cal A}_{\text{I}}={\cal A}_{\text{II}}=0.1$, $L=2$, and $\Omega_0\approx 5$, as a function of the distance $D$. Our framework allows us to fix proper accelerations and therefore study the effect of spatial separation only. In the upper plot in Fig.~\ref{Negativity_vs_D} we consider the range of separations $D$ that are close to zero. When $D$ is positive, the entanglement slowly decreases as a function of $D$, but when $D$ changes  sign, the behavior of the plot changes and we observe rapid oscillations on the increasing curve, as $D$ decreases. Note that in order to satisfy the condition $[\hat{d}_{\text{I}},\hat{d}_{\text{II}}^{(\dagger)}]=0$, we only consider negative values of $D$ such that $\frac{2}{\cal A} + D \gg L$. In Sec.~\ref{compnoise} we have analytically proven that the off-diagonal term of the noise matrix, $N^\pm_{\text{I},\text{II}}$, and hence the logarithmic negativity, is a continuous function at $D=0$, as seen in Eq.~\eqref{N12}. In the lower plot of Fig.~\ref{Negativity_vs_D} we show how ${\cal E_N}$ changes for large, positive values of $D$. As expected, the detected entanglement vanishes for increasing positive separations $D$. We discover a sudden death of entanglement that occurs for a finite distance $D$.

The results plotted in Fig.~\ref{Negativity_fixed_distance} are complementary to the results shown in Fig.~\ref{Negativity_vs_D}. The setting is also symmetric, i.e., the parameters characterizing both output modes are identical. In this example, we fix the spatial separation between the output modes, equal to $\frac{2}{\cal A} + D$, and study the amount of entanglement as a function of the proper acceleration of the output modes. It is clear that in order to fix the distance between the modes, the change in proper acceleration has to be compensated by a change in the separation $D$. The resulting dependence of the detected logarithmic negativity is shown in Fig.~\ref{Negativity_fixed_distance}. In the upper figure, we plot negativity as a function of proper acceleration. We choose the range of the plot such that we can study the transition corresponding to the change of sign of the distance $D$. Again we see that the behavior of ${\cal E}_{{\cal N}}$ changes for these two regimes. The plot exhibits oscillations for $D<0$ and it is smooth for $D>0$. In the lower figure, we plot ${\cal E}_{{\cal N}}$ for a larger range of $D>0$. Once again, we discover a sudden death of entanglement that occurs for finite proper acceleration $\cal A$ and distance $D$. 


Finally we move on to the scenario in which both output modes are accelerated in the same direction, as schematically shown in Fig.~\ref{Para}. Because of the form of Eq. \eqref{N12parallel} it can be shown that for $D=0$ we obtain no entanglement. We have numerically evaluated the logarithmic negativity of the output state of the channel for a range of parameters, and found zero negativity also for $D\neq 0$. This strongly suggests that there is no vacuum entanglement within the parallel accelerations setup.


Let us point out that the framework we have introduced allows one to reliably study the effect of the accelerations of the observers and the distance between them on the observed entanglement. To the best of our knowledge, our results have not been shown previously. We have found that the entanglement witnessed by the accelerated observers in the Minkowski vacuum exhibits an interesting oscillatory behavior, when the two Rindler wedges overlap, i.e., for $D<0$. When the two Rindler wedges are separated by a positive distance, $D>0$, we also find a new phenomenon of sudden death entanglement, similar to the effect reported in \cite{Salton2015}. We believe that the tools presented in this work can be further used to investigate the rich structure of the vacuum field entanglement.

\section{Non-vacuum states\label{secnonvacuum}}
Let us turn our attention towards non-vacuum input Gaussian states. The framework introduced in previous sections allows one to investigate arbitrary Gaussian input states, but we will specialize to the family of two-mode squeezed thermal states. This class of states is sufficiently broad: it covers both separable and entangled states, as well as pure and mixed states, and it is parameterized by only two positive numbers. Consider two thermal states, occupying orthogonal modes $\phi_{\text{I}}$ and $\phi_{\text{II}}$, characterized by the same mean particle numbers $\langle \hat{f}^\dagger_{\text{I}} \hat{f}_{\text{I}}\rangle = \langle \hat{f}^\dagger_{\text{II}} \hat{f}_{\text{II}}\rangle \equiv n$. We will apply a two-mode squeezing operation to these two thermal states, and take the result as an input of the Gaussian channel \eqref{GC}. For the squeezing parameter equal to $r$, the resulting input state is characterized by vanishing vector of first moments, $\vec{X}^{(f)} = \vec{0}$ and the following covariance matrix:
\begin{align}
\label{squeezedtherm}
\sigma^{(f)}&=(1+2n)\left(\begin{array}{cccc}
\cosh 2r & 0 &\sinh 2r & 0\\
0 & \cosh 2r & 0 &-\sinh 2r\\
\sinh 2r & 0 & \cosh 2r & 0\\
0 & -\sinh 2r & 0 & \cosh 2r
 \end{array}\right).
\end{align} 
In the special case of $r=0$ this state reduces to a pair of thermal states, while for $n=0$ it simplifies to a two-mode squeezed vacuum.

In Sec.~\ref{vacent} we have studied the properties of the vacuum entanglement, which boiled down to analyzing the elements of the noise matrix $N$. We have numerically found that the values of $N_{\text{I}}$, $N_{\text{II}}$, and $N_{\text{I,II}}^{\pm}$ were typically several orders of magnitude smaller than unity. For that reason, the contribution to the Gaussian channel \eqref{GC} from these terms can be neglected when the input states \eqref{squeezedtherm} have sufficiently large $n$ or $r$. In this case, the action of the channel \eqref{GC} reduces to:
\begin{align}
\label{GCappr}
\sigma^{(d)}-\openone \approx  M\left(\sigma^{(f)} -\openone\right) M^T,
\end{align}
where $M$ is given by Eq.~\eqref{M}. In this approximation the considered Gaussian channel becomes independent of $D$. Using the relation \eqref{nbeta} we can also neglect the $\beta_\Lambda$ coefficients characterizing matrix $M$, and the dominant contribution to the Fidelity degradation due to acceleration will come from the $\alpha_\Lambda$ coefficients.

Let us now investigate how the input states \eqref{squeezedtherm} are altered by the channel \eqref{GCappr}. A convenient measure that can be used to quantify the discrepancy between the input and the output states of the channel is Uhlmann fidelity ${\cal F}$ \cite{Nielsen}. In our case, when both the input and the output states have vanishing first moments, the Uhlmann fidelity between these states reduces to \cite{Marian2012}:
\begin{align}
\label{fidelity}
{\cal F}\left(\sigma^{(f)},\sigma^{(d)}\right)=
\frac{4}{\sqrt{\lambda}+\sqrt{\gamma}-\sqrt{(\sqrt{\lambda}+\sqrt{\gamma})^2-\delta}},
\end{align}
where:
\begin{align}
\lambda&=\det\left(\openone+i\Sigma\sigma^{(f)}\right)\det\left(\openone+i\Sigma\sigma^{(d)}\right)\nonumber \\
&=16n^2(1+n)^2 \det\left[\openone+i\Sigma+i\Sigma M\left(\sigma^{(f)}-\openone\right)M^T\right],\nonumber \\
\gamma&=\det\left(\openone - \Sigma\sigma^{(f)}\Sigma\sigma^{(d)}\right)\nonumber \\
&=(1+2n)^4\det\left[ \openone + \frac{\sigma^{(f)}}{(1+2n)^2} + M\left(\sigma^{(f)}-\openone\right)M^T\right],\nonumber \\
\delta &=\det\left(\sigma^{(f)}+\sigma^{(d)}\right)\nonumber \\
&=\det\left[ \openone + \sigma^{(f)} + M\left(\sigma^{(f)}-\openone\right)M^T\right],\nonumber \\
\Sigma&=\left(\begin{array}{cccc}
0 & 1 & 0 & 0\\
-1 & 0 & 0 & 0\\
0 & 0 & 0 & 1\\
0 & 0 & -1 & 0
 \end{array}\right).
 \end{align}
In particular, for $n=0$, i.e., when the input state is a pure two-mode squeezed vacuum state, the above expressions simplify significantly. We have $\lambda = 0$, $\gamma = \delta$, and the fidelity reduces to ${\cal F} = \frac{4}{\sqrt{\delta}}$.

\begin{figure}[t]
\includegraphics[width=1\linewidth]{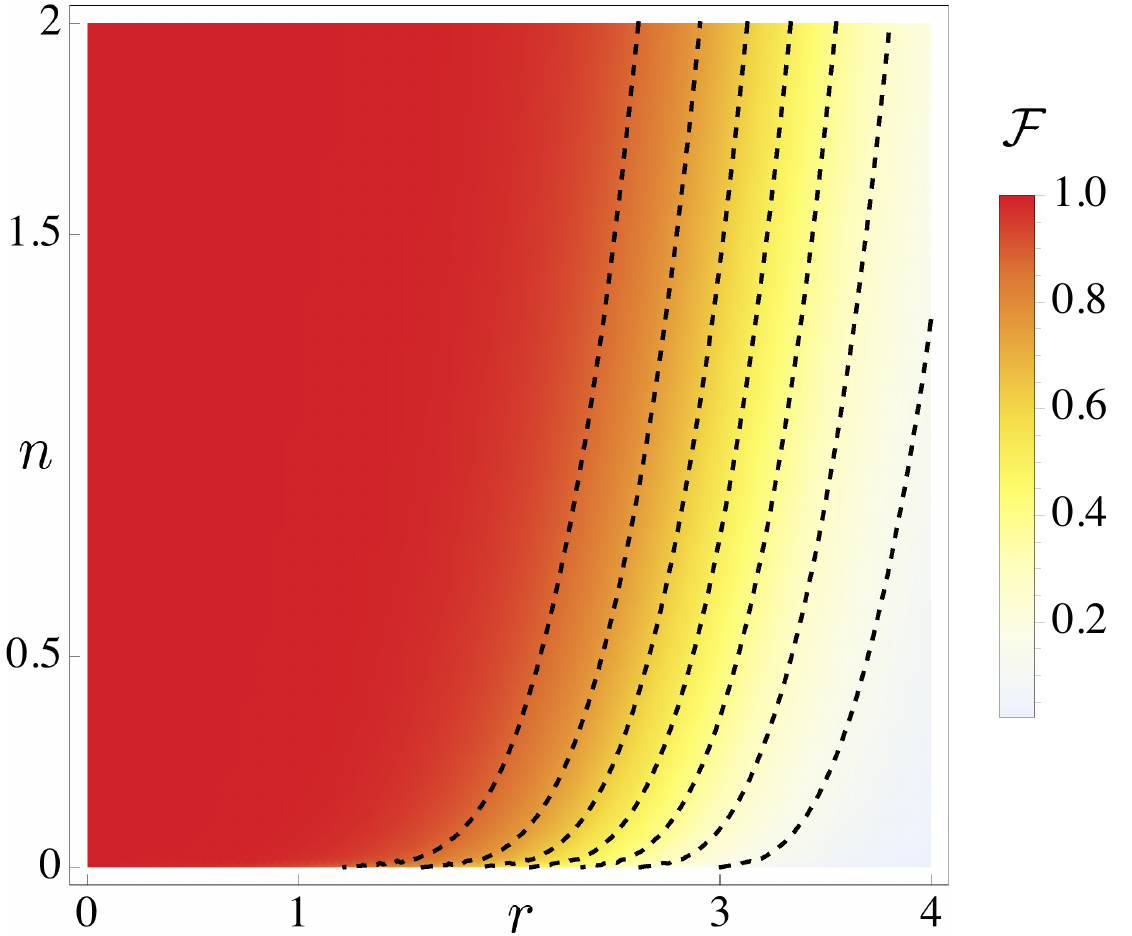}
\caption{\label{Non-Optimum}Fidelity for the input squeezed thermal state as a function of the squeezing parameter $r$ and $n$. We consider two identical counter-accelerated passive modes characterized by ${\cal A}=0.1$, $L=2$, and $\Omega_{0}\approx 5$, with corresponding Bogolyubov coefficients $\ali=\alii=0.985$ and $\bei=\beii=4.51\times 10^{-11}$ and $N_{\text{I}}=N_{\text{II}}=4.82 \times 10^{-10}$, $N_{\text{I,II}}^{\pm}=1.80\times 10^{-9}$. We choose the wedge separation $D=0$ and the field mass $m=0.1$.}
\end{figure}
 
When ${\cal F}$ is close to unity, it indicates that the channel $\eqref{GCappr}$ transforms the input state of the modes $\phi_\Lambda$ into an almost identical state of the modes $\psi_\Lambda$. In this case the channel can be used to reliably send quantum information between the inertial and accelerated observers. When ${\cal F}$ is close to zero, it means that the output state of the channel is almost orthogonal to the input state and effective communication between the observers is impossible.

In Fig.~\ref{Non-Optimum} we plot the fidelity of the output states for the input squeezed thermal states \eqref{squeezedtherm} as a function of their parameters $n$ and $r$. We keep the accelerations and mode parameters fixed at ${\cal A}=0.1$, $L=2$, and $\Omega_{0}\approx 5$. For these values we have computed the parameters characterizing the $M$ matrix \eqref{M}, which boil down to\footnote{Computation of these coefficients is very sensitive to the presence of negative frequency contributions to the modes $\phi_\Lambda$ and $\psi_\Lambda$. For that reason we have introduced a zero-frequency cut-off to the modes \eqref{inputgauss} and \eqref{outputgauss}, although they had very small amounts of negative frequencies to start with. This procedure is applied by replacing the mode overlaps with the following expressions: $\alpha_\Lambda = (\psi_\Lambda,\phi_\Lambda)\to\int\!\!\!\int \text{d}\Omega \text{d}k (\psi_\Lambda,w_{\Lambda\Omega})(w_{\Lambda\Omega},u_k)(u_k,\phi_\Lambda)$, where $(w_{\Lambda\Omega},u_k) = \alpha^{(\Lambda)\star}_{\Omega k}$ given by \eqref{intermsofalpha}. We compute $\beta_\Lambda$ coefficients similarly.}: $\ali=\alii=0.985$ and $\bei=\beii=4.51\times 10^{-11}$. The corresponding parameters specifying the $N$ matrix \eqref{N} were found to be: $N_{\text{I}}=N_{\text{II}}=4.82 \times 10^{-10}$ and $N_{\text{I,II}}^{\pm}=1.80\times 10^{-9}$. The resulting fidelity shown in Fig.~\ref{Non-Optimum} reveals a degradation of fidelity below unity only for highly squeezed thermal states of low temperature. For a higher degree of thermality and less squeezing, the non inertial effects do not play a significant role for the parameters we have chosen. We have also investigated how the fidelity degradation is suppressed when active modes \eqref{activemodes} are considered. For the same set of parameters as those used in Fig.~\ref{Non-Optimum}, we have numerically found that the $\alpha$ coefficients \eqref{alphaactive} are practically equal to unity and consequently the fidelity is nearly perfect.

\begin{figure}[t]
\includegraphics[width=1\linewidth]{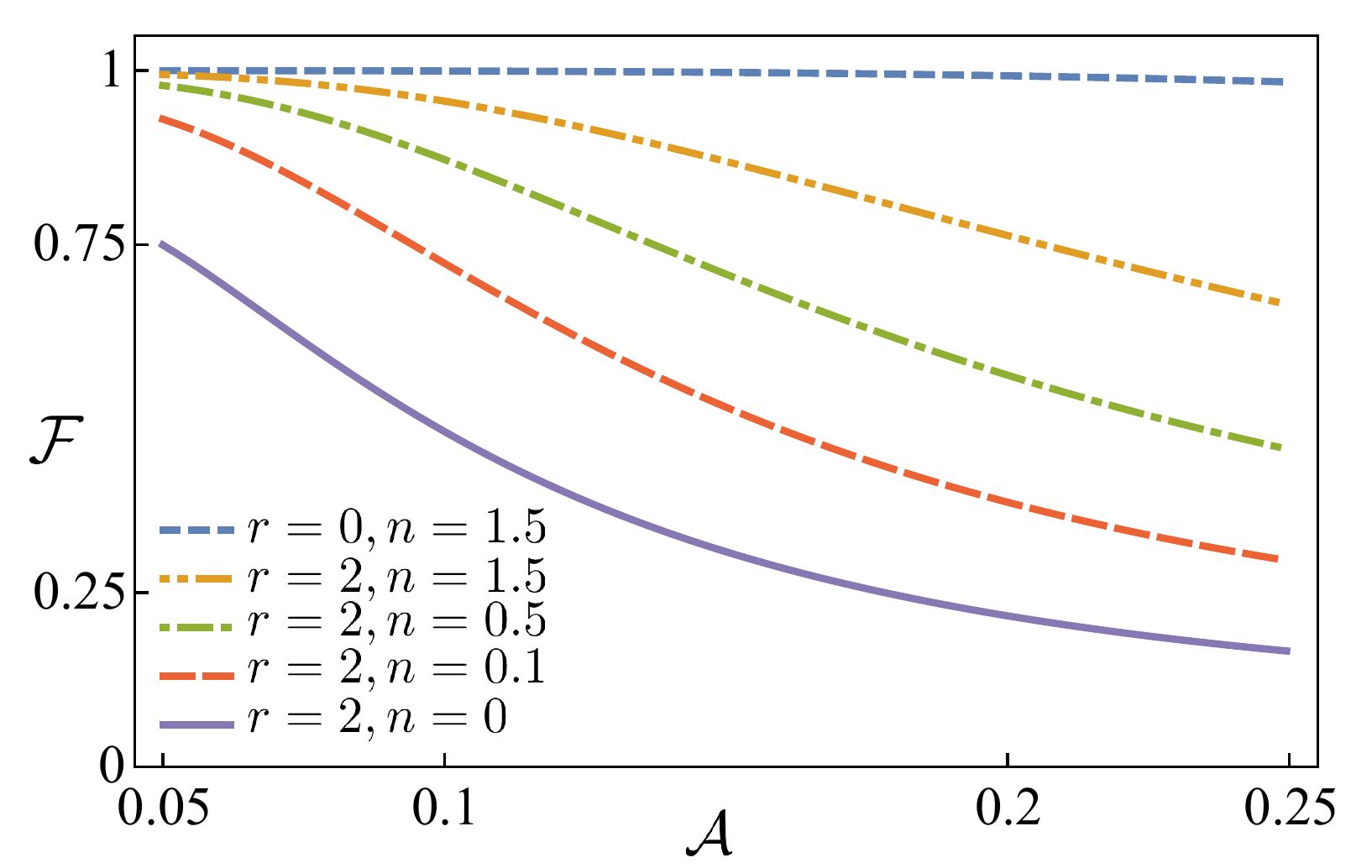}
\caption{\label{Fid-acc}Fidelity for input squeezed thermal states of diverse degrees of squeezing ($r$) and thermality ($n$) as a function of proper acceleration $\cal A$. We consider two identical counter-accelerated passive modes characterized by $L=2$ and $\Omega_{0}\approx 5$, choose the wedge separation $D=0$ and the field mass $m=0.1$.}
\end{figure}

We also study the effect of acceleration on the fidelity of the output states. In Fig.~\eqref{Fid-acc} we fix the mode parameters $L=2$ and $\Omega_{0}\approx 5$ and plot the fidelity of the output state as a function of the acceleration of the output passive modes. We choose a variety of the input states characterized by diverse parameters $n$ and $r$. It is clear from the plots how the increase of the acceleration degrades the fidelity. However the degradation is not an inevitable process as it can be almost completely reduced by choosing the active output modes \eqref{activemodes}.

\newcommand{\Nsm}{N_{\rm sm}}
\newcommand{\Msm}{M_{\rm sm}}
\newcommand{\Nc}{N_{\rm c}}
\newcommand{\Mc}{M_{\rm c}}

\section{Single-mode channel\label{singmode}}
From the two-mode channel constructed in Eq.~\eqref{GC}, a single-mode channel can be constructed by tracing out either mode I or II. For definiteness, let us trace out mode II, which results in the single-mode channel:
\begin{subequations}\label{GCSM}
\begin{align}
\vec{X}^{(d)}_{\rm I} &=  \Msm \vec{X}_{\rm I}^{(f)}, \label{SMchannel1} \\
\sigma^{(d)}_{\rm I} & =  \Msm \sigma^{(f)}_{\rm I} \Msm^T+\Nsm, \label{SMchannel2}
\end{align}
\end{subequations}
where
\begin{subequations}
\begin{align}
\Msm &\equiv \operatorname{Tr}_{\rm II}\!\left(M\right) =
\left( \begin{matrix}
\text{Re}(\ali-\bei) & -\text{Im}(\ali+\bei) \\
\text{Im}(\ali-\bei) & \text{Re}(\ali+\bei) 
\end{matrix}\right), \label{SMM}\\
\Nsm &\equiv \operatorname{Tr}_{\rm II}\! \left(N\right)  =  \left(1+N_{\rm I} \right)\openone - \Msm \Msm^T, \label{SMN}
\end{align}
\end{subequations}
and $N_{\rm I}$ is defined in Eq.~\eqref{N1}; $\vec{X}_{\rm I}^{(i)}$ and $\sigma_{\rm I}^{(i)}$ refer to the first and second statistical moments associated with mode I of the input ($i=f$) and output ($i=d$) modes.

The motivation for studying the simpler single-mode channel is two-fold. First, single-mode Gaussian channels have been fully classified and bounds on their classical and quantum capacities have been found \cite{Holevo2007,Weedbrook2012}. We are thus able to determine the type of single-mode channel our framework results in and to quantify the effect of acceleration has on both classical and quantum channel capacities. Second, from the single-mode channel capacities, we are able to give a (crude) lower bound on  the channel capacity of the two-mode channel discussed in the previous sections. Explicitly the channel capacity of the two-mode channel is at least twice the channel capacity of the single-mode channel.

Two Gaussian channels are said to be equivalent if there exists a unitary operation on the input and a unitary operation on the output of the channel that maps one channel into the other. Physically, these operations correspond to pre-processing by the sender, in our case the inertial observer, by applying a Gaussian unitary represented by the symplectic matrix $S_{\rm in}$ to the input state:
\begin{align}
\vec{X}^{(f)}_{\text I} &\to  S_{\rm in} \vec{X}^{(f)}_{\text I},  &
\sigma^{(f)}_{\text I} &\to S_{\rm in} \sigma^{(f)}_{\text I}  S_{\rm in}^T,
\end{align}
and post-processing by the receiver, the uniformly accelerating observer, by applying a Gaussian unitary $S_{\rm out}$ to the output state:
\begin{align}
\vec{X}^{(d)}_{\text I} &\to  S_{\rm out} \vec{X}^{(d)}_{\text I},  &
\sigma^{(d)}_{\text I} &\to S_{\rm out} \sigma^{(d)}_{\text I}  S_{\rm out}^T.
\end{align}
Through pre- and post-processing the single-mode channel above may be brought to its canonical form and classified accordingly \cite{Holevo2007,Weedbrook2012}. 

The pre- and post-processing operations described above can be equivalently seen, with the help of Eqs.~\eqref{SMchannel1} and \eqref{SMchannel2},  as transforming the matrices $\Msm$ and $\Nsm$ to a canonical form:
\begin{subequations}
\begin{align}
\Msm \to  \Mc &= S_{\rm out} \Msm S_{\rm in}, \\
\Nsm \to   \Nc &= S_{\rm out} \Nsm S_{\rm out}^T.
\end{align}
\end{subequations}

The explicit form of $\Mc$ and $\Nc$ depend on three quantities which are left invariant under the action of the pre- and post-processing operations $S_{\rm in}$ and $S_{\rm out}$. These invariants are the rank of the channel $\min \left\{\operatorname{rank}(\Msm), \operatorname{rank} (\Nsm)  \right\}$, the generalized transmissivity $\tau\equiv \det \Msm $, and the thermal number:
\begin{align}
\bar{n} \equiv
\begin{cases}
\sqrt{\det \Nsm}, &\mbox{for } \tau=1, \\ 
\frac{1}{2\abs{1-\tau}}\sqrt{\det \Nsm} - \frac{1}{2}, & \mbox{for } \tau \neq 1. 
\end{cases} \label{thermalnumber}
\end{align}

By inspection of Eqs.~\eqref{SMM} and \eqref{SMN}, the rank of the channel is determined to be $2$. From Eq.~\eqref{SMM} we find the transmissivity to be:
\begin{align}
\tau = \abs{\alpha_{\rm I}}^2- \abs{\beta_{\rm I}}^2.
\end{align}
By definition both $\abs{\alpha_{\rm I}}^2$ and $\abs{\beta_{\rm I}}^2$ are in the interval $[0,1]$, and thus the transmissivity $\tau$ lies in the interval $[-1,1]$. From Eq.~\eqref{SMN} the determinant of $\Nsm$ may be directly calculated:
\begin{align}
\det \Nsm = \Big[ \abs{\alpha_{\rm I}}^2 + \abs{\beta_{\rm I}}^2   - \left(1+ N_{\rm I}\right) \Big]^2 - 4 \abs{\alpha_{\rm I}}^2 \abs{\beta_{\rm I}}^2,
\end{align}
and substituted into Eq.~\eqref{thermalnumber} to determine the thermal number associated with the channel.

Having determined the rank and the transmissivity of the channel in Eq.~\eqref{GCSM}, we may identify it as a lossy channel, and express it in its canonical form \cite{Holevo2007}:
\begin{subequations}
\begin{align}
\Mc &= \sqrt{\tau} \openone, \\
\Nc &= \left(1-\tau\right) \left( 2 \bar{n}+1\right) \openone. \label{Nc}
\end{align}
\end{subequations}

\begin{figure}[t]
\includegraphics[width=1\linewidth]{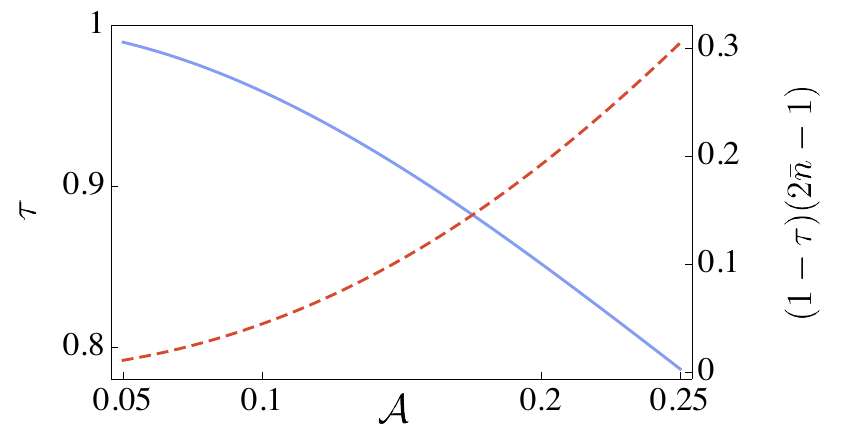}
\caption{\label{SMplot} The transmissivity $\tau$ (solid line) and the coefficient $(1-\tau)(2\bar{n}+1)$ appearing in Eq.~\eqref{Nc} (dashed line) associated with the single-mode channel are plotted as a function of the {acceleration of the non-inertial mode for $L=2$, $\Omega_0\approx 5$, and $m=0.1$.}}
\end{figure}

For the choice of mode functions in Eqs.~\eqref{inputgauss} and \eqref{outputgauss}, both the transmissivity of the channel $\tau$, quantifying the attenuation of the channel, and the coefficient appearing in front of $\openone$ in Eq.~\eqref{Nc}, quantifying the amount of noise in the channel, are plotted as a function of the acceleration of the output mode in Fig.~\ref{SMplot}. We find that the single-mode channel is more attenuating and noisy as the acceleration of the output mode increases.

For a lossy channel the classical capacity $C$ is known to be bound from below by \cite{Lupo2011}:
\begin{align}
C \geq g\Big(2 \tau (\bar{m} - \bar{n}) + 2 \bar{n} +1 \Big) - g\Big( 2 \bar{n} (1-\tau) \Big), \label{ClassicalCap}
\end{align}
where 
$\bar{m}$ is the mean number of particles in the input state \cite{Holevo1999}, and we have introduced the function:
\begin{align}
g(x) \equiv \left(\frac{x+1}{2}\right) \log \left(\frac{x+1}{2}\right) - \left(\frac{x-1}{2}\right) \log \left(\frac{x-1}{2}\right);
\end{align}
this bound is believed to be tight \cite{Weedbrook2012}.

A lower bound on the quantum capacity $Q$ of the channel may also be given \cite{Holevo2001,Pirandola2009}:
\begin{align}
Q \geq \max\left[ 0, \log \left|\frac{\tau}{1-\tau}\right| - g(\nu)\right]. \label{QuantumCap}
\end{align}

The lower bounds on the classical and quantum capacities are plotted as a function of acceleration in Fig. \ref{capacities}. As expected, based on the behavior of the transmissivity and noise present in the channel illustrated in Fig.~\ref{SMplot}, both bounds on the capacities decrease as a function of the acceleration of the {non-inertial mode.

\begin{figure}[t]
\includegraphics[width=1\linewidth]{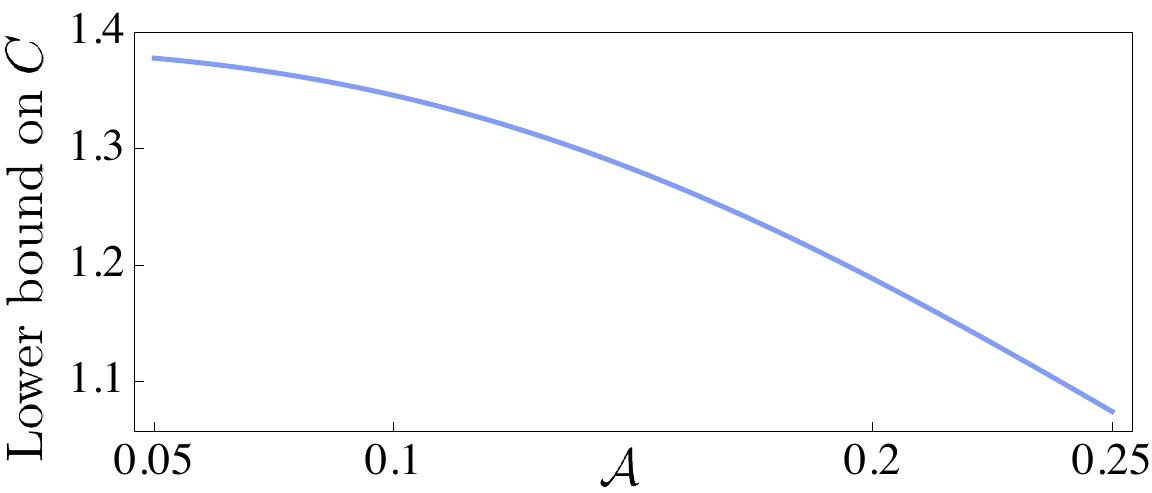} 
\\ \ 
\\
\includegraphics[width=1\linewidth]{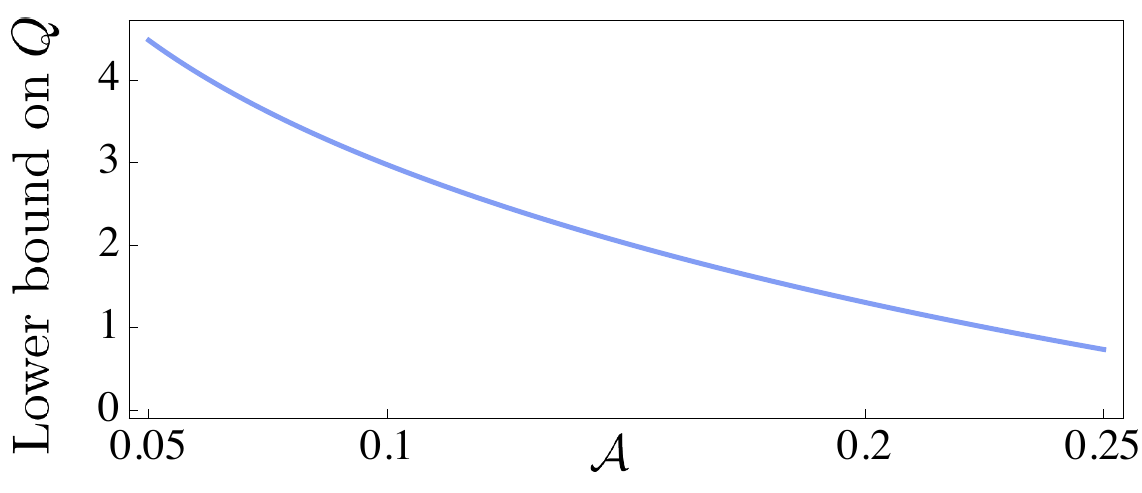}
\caption{\label{capacities} Both the lower bound on the classical capacity for $\bar{m}=1$ given in Eq.~\eqref{ClassicalCap} (upper plot), and the lower bound on the quantum capacity given in Eq.~\eqref{QuantumCap} (lower plot) are plotted as a function of acceleration for for $L=2$, $\Omega_0\approx 5$, and $m=0.1$.}
\end{figure}

\section{Discussions and outlook\label{concl}}

In this paper, we studied how an arbitrary Gaussian state of two localized inertial wave packets of a massive real scalar field is described by a pair of uniformly accelerated observers. We demonstrated that such a change of reference frame can be formulated as the action of a Gaussian quantum channel, which we derived analytically. Using this channel, we analyzed several different scenarios. These include the situations wherein the Rindler wedges corresponding to counter-accelerating observers are separated by an arbitrary negative or positive distance as shown in Figs.~\ref{Dp} and \ref{Dn}, as well as the scenario in which the observers are co-accelerating, as depicted in Fig.~\ref{Para}. 

We analyzed the amount of entanglement that can be extracted from the Minkowski vacuum in all the above-mentioned scenarios. As expected, we observed that vacuum entanglement is an increasing function of proper accelerations when the two Rindler wedges have a common apex (Fig.~\ref{D0_vs_Accelerations}). In this case, we also found that to extract most vacuum entanglement from the field, the width of the wave packets and also their central frequency has to be as small as possible (Fig.~\ref{Negativity_vs_LOmega}).  

We found very sensitive and non-monotonous dependence of the vacuum entanglement as a function of separation for the $D<0$ case. Furthermore, we observed sudden death of entanglement in two situations. The first one was sudden death of vacuum entanglement as a function of the distance between the accelerated modes while their proper accelerations were kept unchanged (Fig.~\ref{Negativity_vs_D}). The second one was sudden death of entanglement as a function of the proper acceleration when the distance between them was fixed (Fig.~\ref{Negativity_fixed_distance}). Moreover, we analytically proved that the transition between the two cases of $D>0$ and $D<0$ is continuous for both counter-accelerating and co-accelerating observers.\\

Entanglement sudden death is a curious effect that is usually relegated to early stage disentanglement of quantum systems due to decoherence effects such as amplitude or phase dephasing or a mixture of both \cite{Yu2009}. This phenomena is generally studied in the context of open quantum systems, where entanglement suddenly vanishes as the system evolves in time. However the two sudden deaths of entanglement that we have reported in this paper refer to the disappearance of entanglement between the two modes of a massive quantum field, namely the modes $\text{I}$ and $\text{II}$, as a function of the proper acceleration of the accelerating observers and the distance between them. 

In open quantum systems, it was shown that quantum error correction can delay this unfavourable effect in some cases \cite{Sainz2008}. As a future line of research, it would be interesting to investigate if quantum error correction methods can be devised to delay or even avoid the reported  disentanglement of the vacuum state of the quantum field. To do so
would necessitate  reducing the two terms $N_{\text{I}}$ and $N_{\text{II}}$ while keeping the terms $N^\pm_{\text{I,II}}$ as large as possible. Also, as (Fig.~\ref{Negativity_vs_LOmega}) shows, highly localized wavepackets with smaller central frequencies increase the observed vacuum entanglement for the case of  $D=0$. Therefore, another possible approach is to perform a similar optimization over the central frequency of the wavepacket $\Omega_0$ and its width $L$ as a function of the distance $D$ and the proper acceleration $\cal A$.

We have also analyzed the family of two-mode squeezed thermal states as the input of the channel. We have investigated the fidelity of the output states as a function of the purity and non-locality of the input states, as well as proper acceleration of the observer at the output.

Finally, we analyzed the situation in which only one accelerated party observes the output of the channel. In such a scenario we were able to evaluate lower bounds on classical and quantum capacities of the corresponding single-mode channel.

Our framework can readily be applied to quantum information protocols in which the effect of acceleration or gravity cannot be ignored. As a future line of research, we are interested in employing our framework to study the effect of acceleration on communication between two relativistic observers without a shared reference frame \cite{Mehdi2015,Agata2015} as well as continuous variable teleportation \cite{Braunstein1998,Nico2013}. We are also interested in investigating analogous effects in other types of quantum fields such as massless (that cannot be obtained by taking $m\to 0$ limit of the results of this paper; see footnote~\ref{mfoot} in Sec.~\ref{compnoise}) \cite{lorek} and fermionic fields \cite{richter}.     

\acknowledgments
We thank Jorma Louko, Rafa{\l} Demkowicz-Dobrza\'{n}ski Jason Doukas, Filip Kia{\l}ka, Zehua Tian and Piotr Grochowski for useful discussions and comments. This work was supported by National Science Centre, Sonata BIS Grant No. DEC-2012/07/E/ST2/01402.

\appendix
\section{Computing the noise matrix $N$\label{computingnoise}}
In this Appendix we compute the output state of the channel for the input Minkowski vacuum, i.e., $\sigma_{\text{vac}}^{(d)}$. We use this state to determine the noise matrix, as can be seen in Eq.~\eqref{N}. We first focus on the case of $D=0$, and then continue the analysis for a more challenging case of $D\neq 0$ and eventually consider the scenario, wherein both output modes are accelerated in the same direction.
\subsection{When Rindler wedges have a common apex ($D=0$)}\label{D0}
First, we consider the simplest case of $D=0$, i.e., when the two Rindler wedges have a common apex. Let us begin by computing the first element of the covariance matrix $\sigma_{\text{vac}}^{(d)}$. Using the definition of covariance matrix \eqref{CM} we have:$ \left(\sigma^{(d)}_{\text{vac}}\right)_{11} =
{}_\text{M}\bra{0}\left( \hat{d}_\text{I}+\hat{d}\d_\text{I} \right)
\left( \hat{d}_\text{I}+\hat{d}\d_\text{I} \right)
\ket{0}_{\text{M}}.$ After substituting the decomposition of $\hat{d}_\text{I}$ and $\hat{d}_\text{I}^{\dag}$ in terms of Rindler operators, as given in \eqref{detrinbogoI}, we get:
\begin{align}
\label{sigma11}
\left(\sigma^{(d)}_{\text{vac}}\right)_{11} =&\int\!\!\!\!\int\mathrm{d}\Omega\,\mathrm{d}\Xi\,{}_\text{M}\bra{0}
\left[ (\psi_\text{I},\wi[\Omega])\bi[\Omega] + (\psi_\text{I},\wi[\Omega])^\star\bi[\Omega]\d\right]\nonumber\\
&\times
\left[  (\psi_\text{I},\wi[\Xi])\bi[\Xi] + (\psi_\text{I},\wi[\Xi])^\star\bi[\Xi]\d\right]
\ket{0}_{\text{M}}.
\end{align}
In order to proceed further we need to evaluate the expectation values of the quadratic forms of the Rindler operators on the Minkowski vacuum state that appear above. For $D=0$ the Bogolyubov transformation between the Minkowski modes and the Rindler modes~\eqref{minrindec} is given by: 
\begin{align}\label{rinmind0}
\hat{b}_{\Lambda\Omega}&=\int\mathrm{d}k\,\left(\alpha_{\Omega k}^{(\Lambda)\star}\hat{a}_k-\beta_{\Omega k}^{(\Lambda)\star}\hat{a}_k\d\right),
\end{align}
where $\Lambda\in\{\text{I}, \text{II}\}$ and the Bogolyubov coefficients $\alpha_{\Omega k}^{(\Lambda)} = (u_k,w_{\Lambda\Omega})$ and $\beta_{\Omega k}^{(\Lambda)} = -(u_k^\star,w_{\Lambda\Omega})$ are given by \cite{Crispino2008}: 
\begin{align}\label{intermsofalpha}
\alpha_{\Omega k}^{(\text{I})} &=\frac{1}{\sqrt{4\pi\omega_k a \sinh\left(\frac{\pi\Omega}{a}\right)}}
\left(\frac{\omega_k+k}{\omega_k-k}\right)^{-i\frac{\Omega}{2a}}e^{\frac{\pi\Omega}{2a}},
\nonumber \\
\alpha_{\Omega k}^{(\text{II})} &= {\alpha_{\Omega k}^{(\text{I})}}^\star,\nonumber \\
\beta_{\Omega k}^{(\text{I})} &= -e^{-\frac{\pi\Omega}{a}}\alpha_{\Omega k}^{(\text{I})},\nonumber \\
\beta_{\Omega k}^{(\text{II})} &=-e^{-\frac{\pi\Omega}{a}} {\alpha_{\Omega k}^{(\text{I})}}^\star.
\end{align}
In order for the transformation to preserve commutation relations, the following Bogolyubov identities must hold \cite{BirrelDavies}:
\begin{align}\label{Bogoiden}
\int \mathrm{d}k \big(\alpha_{\Omega k}^{(\Lambda)} \alpha_{\Xi k}^{(\Lambda)\star} - \beta_{\Omega k}^{(\Lambda)} \beta_{\Xi k}^{(\Lambda)\star} \big)&=\delta(\Omega-\Xi),\\
\int \mathrm{d}k \big(\alpha_{\Omega k}^{(\Lambda)} \beta_{\Xi k}^{(\Lambda)} - \beta_{\Omega k}^{(\Lambda)} \alpha_{\Xi k}^{(\Lambda)} \big)&=0.
\end{align}
Using the relation \eqref{intermsofalpha} between coefficients $\alpha_{\Omega k}^{(\Lambda)}$ and $\beta_{\Omega k}^{(\Lambda)}$, i.e., $
\beta_{\Omega k}^{(\Lambda)}=-e^{-\frac{\pi\Omega}{a}}\alpha_{\Omega k}^{ (\Lambda) }$,
we can rewrite the Bogolyubov identities as:
\begin{align}\label{BIR}
\int \mathrm{d}k\,\alpha_{\Omega k}^{(\Lambda)} \alpha_{\Xi k}^{(\Lambda)\star}&=
\frac{\delta(\Omega-\Xi)}{1-e^{-\frac{2\pi\Omega}{a}}},\nonumber\\
\int \mathrm{d}k\,\alpha_{\Omega k}^{(\Lambda)} \alpha_{\Xi k}^{(\Lambda)}&=0.
\end{align}
Then using Eqs.~\eqref{rinmind0}, \eqref{intermsofalpha}, and \eqref{BIR} we can write:
\begin{align}
{}_\text{M}\bra{0} \bl[\Omega]\bl[\Xi] \ket{0}_{\text{M}} &= {}_\text{M}\bra{0} \bl[\Omega]\d\bl[\Xi]\d \ket{0}_{\text{M}}\nonumber\\
&\propto \int \mathrm{d}k\,\alpha_{\Omega k}^{(\Lambda)} \alpha_{\Xi k}^{(\Lambda)}=0, \label{quadr1}\\
{}_\text{M}\bra{0} \bl[\Omega]\bl[\Xi]\d \ket{0}_{\text{M}}&=\int \mathrm{d}k\,\alpha_{\Omega k}^{(\Lambda)\star} \alpha_{\Xi k}^{(\Lambda)}=\frac{\delta(\Omega-\Xi)}{1-e^{-\frac{2\pi\Omega}{a}}}\label{quadr2}.
\end{align}
We also find: 
\begin{align}
\label{bed}
{}_\text{M}\bra{0} \bl[\Omega]\d\bl[\Xi] \ket{0}_{\text{M}}&=e^{-\frac{\pi(\Omega+\Xi)}{a}}\int \mathrm{d}k \alpha_{\Omega k}^{(\Lambda)}
\alpha_{\Xi k}^{(\Lambda)\star}=\frac{\delta(\Omega-\Xi)}{e^{\frac{2\pi\Omega}{a}}-1}.
\end{align}
The above relation shows that the spectrum of the average Rindler particle number in Minkowski vacuum is given by the Bose-Einstein distribution. At this point, it is worth pointing out that the parameter $a$ appearing in the above expressions should not be confused with the proper acceleration of the observer. Sec.~\ref{ain} is devoted to the in-depth discussion of this fact.

Finally, we use the relations \eqref{intermsofalpha} between Bogolyubov coefficients, namely $\alpha_{\Omega k}^{(\text{II})}=\alpha_{\Omega k}^{(\text{I})*}$ and 
$\beta_{\Omega k}^{(\text{II})}=-\alpha_{\Omega k}^{(\text{I})*}e^{-\frac{\pi\Omega}{a}}$,   
to compute the remaining quadratic expectation values:
\begin{align}
{}_\text{M}\bra{0} \bi[\Omega]\bii[\Xi] \ket{0}_{\text{M}} &={}_\text{M}\bra{0} \bi[\Omega]\d\bii[\Xi]\d \ket{0}_{\text{M}}\nonumber \\
&=e^{-\pi\Omega/a}\int \mathrm{d}k\, \alpha_{\Omega k}^{(\text{I})}\alpha_{\Xi k}^{(\text{I})*}=\frac{\delta(\Omega-\Xi)}{2\sinh\left(\frac{\pi\Omega}{a}\right)}
\label{quadr3}
\end{align}
and
\begin{align}
{}_\text{M}\bra{0} \bi[\Omega]\d\bii[\Xi] \ket{0}_{\text{M}} &= {}_\text{M}\bra{0} \bi[\Omega]\bii[\Xi]\d \ket{0}_{\text{M}} \propto  \int \mathrm{d}k \, \alpha_{\Omega k}^{(\text{I})} \alpha_{\Xi k}^{(\text{I})}=0.
\label{quadr4}
\end{align}

We can now proceed by substituting the obtained expectation values into Eq.~\eqref{sigma11}. Using Eqs.~\eqref{quadr1}, \eqref{quadr2}, and \eqref{bed} in the Eq.~\eqref{sigma11} and assuming normalization of the mode $\psi_{\text{I}}$ we find:
\begin{align}
\left(\sigma^{(d)}_{\text{vac}}\right)_{11}= 1+\int \mathrm{d}\Omega\, \frac{|(\psii,\wi[\Omega])|^2}{\sinh\left(\frac{\pi \Omega}{a}\right)}e^{-\frac{\pi \Omega}{a}}. 
\end{align}
In evaluating the covariance matrix element $\left(\sigma^{(d)}_{\text{vac}}\right)_{22}={}_\text{M}\bra{0}\left( \hat{d}_\text{I}-\hat{d}_\text{I}\d\right)
\left( \hat{d}\d_\text{I}-\hat{d}_\text{I} \right)
\ket{0}_{\text{M}}$, we only need to change the sign of two of the terms in \eqref{sigma11}. Since these terms vanish, we have: 
\begin{align}
\left(\sigma^{(d)}_{\text{vac}}\right)_{22}=1+\int \mathrm{d}\Omega\, \frac{|(\psii,\wi[\Omega])|^2}{\sinh\left(\frac{\pi \Omega}{a}\right)}e^{-\frac{\pi \Omega}{a}} = \left(\sigma^{(d)}_{\text{vac}}\right)_{11}.
\end{align}
The remaining diagonal elements can be similarly computed by replacing the subscript I with II everywhere in Eq.~\eqref{sigma11}, which gives us: 
\begin{align}
\left(\sigma^{(d)}_{\text{vac}}\right)_{33}=\left(\sigma^{(d)}_{\text{vac}}\right)_{44}=1+\int \mathrm{d}\Omega\, \frac{|(\psiii,\wii[\Omega])|^2}{\sinh\left(\frac{\pi \Omega}{a}\right)}e^{-\frac{\pi \Omega}{a}}.
\end{align}

We continue by computing the off-diagonal elements of the upper left block of the covariance matrix. Using the definition \eqref{CM} we have: $\left(\sigma^{(d)}_{\text{vac}}\right)_{12}=\frac{1}{2i}{}_\text{M}\bra{0}\left\{\hat{d}_{\text{I}}+\hat{d}\d_{\text{I}},\hat{d}_{\text{I}}-\hat{d}\d_{\text{I}}\right\}\ket{0}_{\text{M}}=
2\,\Im \,{}_\text{M}\bra{0}{\hat{d}_\text{I}}^2\ket{0}_{\text{M}}$. Then we express the operator $\hat{d}_{\text{I}}$ in terms of Rindler operators as given in Eq.~\eqref{detrinbogoI} and use Eq.~\eqref{quadr1} to get: 
\begin{align}\label{sigma12}
\left(\sigma^{(d)}_{\text{vac}}\right)_{12}\!\!\!\!\!&=2\,\Im\int\!\!\!\!\int\mathrm{d}\Omega\,\mathrm{d}\Xi\, (\psi_\text{I},\wi[\Omega]) (\psi_\text{I},\wi[\Xi]){}_\text{M}\bra{0}
\bi[\Omega]\bi[\Xi] \ket{0}_{\text{M}}\nonumber\\&=0 = \left(\sigma^{(d)}_{\text{vac}}\right)_{21}.
\end{align}
Note that the off-diagonal elements of the lower right block of the covariance matrix can be simply computed by replacing the subscripts I with II in Eq.~\eqref{sigma12} which gives us $\left(\sigma^{(d)}_{\text{vac}}\right)_{34}\!\!\!=\left(\sigma^{(d)}_{\text{vac}}\right)_{43}\!\!\!=0$.  To obtain the upper right block of the covariance matrix,
we substitute the decomposition of $\hat{d}_\text{I}$ and $\hat{d}_\text{II}$ in terms of Rindler operators \eqref{detrinbogo} and use the fact that $[\hat{d}_\text{I},\hat{d}_\text{II}^{(\dag)}]=0$.  For $\left(\sigma^{(d)}_{\text{vac}}\right)_{13}\!\!\!\!=\!\!{}_\text{M}\bra{0}\left( \hat{d}_\text{I}+\hat{d}\d_\text{I}\right)
\left( \hat{d}_\text{II}+\hat{d}\d_\text{II} \right)
\ket{0}_{\text{M}}$, we arrive at:
\begin{align}\label{sigma13}
\left(\sigma^{(d)}_{\text{vac}}\right)_{13} =&
\int\!\!\!\!\int\mathrm{d}\Omega\,\mathrm{d}\Xi\,{}_\text{M}\bra{0}
\left[ (\psi_\text{I},\wi[\Omega])\bi[\Omega] + (\psi_\text{I},\wi[\Omega])^\star\bi[\Omega]\d \right]\nonumber\\
&\times\left[ (\psi_\text{II},\wii[\Xi])\bii[\Xi] + (\psi_\text{II},\wii[\Xi])^\star\bii[\Xi]\d \right]
\ket{0}_{\text{M}}.
\end{align}
Upon substitution of the expectation values \eqref{quadr3} and \eqref{quadr4} into Eq.~\eqref{sigma13} we obtain: 
\be
\left(\sigma^{(d)}_{\text{vac}}\right)_{13}\!\!\!\!=\Re\int\mathrm{d}\Omega\,\frac{
(\psi_\text{I},\wi[\Omega])(\psi_\text{II},\wii[\Omega])}{\sinh\left(\frac{\pi\Omega}{a}\right)}.
\ee
The remaining diagonal element of the upper right block, $\left(\sigma^{(d)}_{\text{vac}}\right)_{24}\!\!\!\!=\!\!{}_\text{M}\bra{0}\left( \hat{d}_\text{I}-\hat{d}_\text{I}\d\right)
\left( \hat{d}\d_\text{II}-\hat{d}_\text{II} \right)
\ket{0}_{\text{M}}$, can be computed analogously and the result reads as:
 \begin{align}\label{sigma24}
&\left(\sigma^{(d)}_{\text{vac}}\right)_{24}\!\!\!\!=-\Re\int\mathrm{d}\Omega\,\frac{(\psi_\text{I},\wi[\Omega])(\psi_\text{II},\wii[\Omega])}{\sinh\left(\frac{\pi \Omega}{a}\right)}.
\end{align}

The last two elements of the covariance matrix 
$\left(\sigma^{(d)}_{\text{vac}}\right)_{14}\!\!\!\!={}_\text{M}\bra{0}i\left( \hat{d}_\text{I}+\hat{d}_\text{I}\d\right)
\left( \hat{d}\d_\text{II}-\hat{d}_\text{II} \right)
\ket{0}_{\text{M}}$ and $\left(\sigma^{(d)}_{\text{vac}}\right)_{23}\!\!\!\!=\!\!{}_\text{M}\bra{0}i\left( \hat{d}_\text{I}\d-\hat{d}_\text{I}\right)
\left( \hat{d}\d_\text{II}+\hat{d}_\text{II} \right)
\ket{0}_{\text{M}}$ can be computed by first multiplying the whole summation in Eq.~\eqref{sigma13} with an imaginary factor $i$ and then changing the sign of some of the terms according to the definition of covariance matrix elements in Eq.~\eqref{CM}. This way we find these two elements to be:
\begin{align} 
&\left(\sigma^{(d)}_{\text{vac}}\right)_{14}=\left(\sigma^{(d)}_{\text{vac}}\right)_{23}=\Im\int\mathrm{d}\Omega\,\frac{(\psi_\text{I},\wi[\Omega])(\psi_\text{II},\wii[\Omega])}{\sinh\left(\frac{\pi \Omega}{a}\right)}.\nonumber\\ 
\end{align}
Note that the lower left block of the covariance matrix is equal to the transpose of the upper right block, therefore we have completed the computation of the covariance matrix $\sigma^{(d)}_{\text{vac}}$. 

The above elements of the covariance matrix of the output state, $\sigma^{(d)}_{\text{vac}}$ can be now plugged into Eq.~\eqref{N} to obtain the results given by Eqs.~\eqref{noise}, \eqref{Ns}, and \eqref{N12}.

Before we proceed with considering the case of $D\neq 0$, let us derive the relation \eqref{nbeta} between the matrix elements $N_\Lambda$ and $\beta_\Lambda$. We have by definition:
\begin{equation}
|\beta_\Lambda|^2 = |-(\psi_\Lambda,\phi^\star_\Lambda)|^2 = \left| \int\text{d}k(\psi_\Lambda,u^\star_k)(u^\star_k,\phi^\star_\Lambda) \right|^2.
\end{equation}
Using Cauchy-Schwarz inequality $|\int fg |^2 \leq \int |f|^2 \int |g|^2$ we obtain:
\begin{align}
|\beta_\Lambda|^2 &\leq \int\text{d}k\left|(\psi_\Lambda,u^\star_k)\right|^2 \nonumber \\
&= \int\text{d}k\left|\int\text{d}\Omega(\psi_\Lambda,w_\Omega)(w_\Omega,u^\star_k) \right|^2\nonumber \\
&=\int\!\!\!\!\int\!\!\!\!\int\text{d}k\,\text{d}\Omega\,\text{d}\Xi\,(\psi_\Lambda,w_\Omega)(\psi_\Lambda,w_\Xi)^\star\beta^{(\Lambda)\star}_{\Omega k}\beta^{(\Lambda)}_{\Xi k},
\end{align}
where we have introduced Bogolyubov coefficients $\beta^{(\Lambda)}_{\Omega k}$ defined in \eqref{intermsofalpha}. Then applying a relation between $\beta^{(\Lambda)}_{\Omega k}$ and $\alpha^{(\Lambda)}_{\Omega k}$ coefficients \eqref{intermsofalpha} and the first Bogolyubov identity \eqref{BIR} we finally arrive at:
\begin{equation}
|\beta_\Lambda|^2 \leq \int\text{d}\Omega\,\frac{|(\psi_\Lambda,w_\Omega)|^2}{1-e^{-\frac{2\pi\Omega}{a}}}e^{-\frac{2\pi\Omega}{a}} = \frac{1}{2} N_\Lambda,
\end{equation}
which is equivalent to Eq.~\eqref{nbeta}.

\subsection{When Rindler wedges do not have a common apex ($D\neq 0$)\label{Dnonzero}}
Our computation of the noise matrix $N$ for the case of $D\neq 0$ proceeds in a very similar fashion to the case of $D=0$. For the latter case we used the Bogolyubov identities to compute the expectation values of Rindler operators, however when $D \neq 0$ another method must be employed. Note that, similar to the case $D=0$, we assume that the operators in Eqs.~\eqref{detrinbogoI} and \eqref{detrinbogoII} are constructed in such a way that they commute, i.e., $\left[\hat{d}_\text{I},\hat{d}^{(\dagger)}_\text{II}\right] = 0$. When $D>0$, this property of the operators is trivially satisfied, however for $D<0$, this assumption limits our freedom in choosing $\psii$ and $\psiii$. In order to secure the above condition we assume that the modes $\psii$ and $\psiii$ are chosen such that their supports do not overlap, which is a sufficient condition for the corresponding operators to commute.

In order to approach the case of $D\neq 0$, we first analyze how the Minkowski-Rindler Bogolyubov coefficients are modified when $D\neq 0$ as compared to the $D=0$ case. For $D=0$ the Bogolyubov transformation between Minkowski modes \eqref{Minkowskimodes} and Rindler modes \eqref{rinmodes} is given by Eq.~\eqref{rinmind0} and the Bogolyubov coefficient are given by Eqs.~\eqref{intermsofalpha}.

Let us investigate how these Bogolyubov coefficients are modified when $D>0$. As depicted in Figs.~\ref{Dp} and \ref{Dn}, region~I is shifted by $\frac{D}{2}$, therefore to modify the Bogolyubov coefficient $\alpha_{\Omega k}^{(\text{I})}$ we can equivalently shift the Minkowski plane-wave solutions \eqref{Minkowskimodes} in the opposite direction: $u_k(x,t)$ is replaced by $\tilde{u}_k(x,t)\equiv u_k(x+\frac{D}{2},t)=e^{i\frac{D}{2}k}u_k(x,t)$. Using the anti-linearity of the scalar product we find the modified $\alpha_{\Omega k}^{\text{(I)}}$ coefficient to be: 
\begin{align}\label{modbogo}
(\tilde{u}_k,w_{\text{I}\Omega})
&=e^{-i \frac{D}{2}k}(u_k,w_{\text{I}\Omega})=e^{-i \frac{D}{2}k}\alpha_{\Omega k}^{\text{(I)}}.
\end{align}
Similarly one can check that the rest of the Bogolyubov coefficients are modified as: 
\begin{align}\label{modifiedbogos}
\alpha_{\Omega k}^{\text{(I)}}& \to e^{-i \frac{D}{2}k}\alpha_{\Omega k}^{\text{(I)}}, &\beta_{\Omega k}^{\text{(I)}} &\to e^{i \frac{D}{2}k}\beta_{\Omega k}^{\text{(I)}},\nonumber\\
\alpha_{\Omega k}^{\text{(II)}}& \to e^{i \frac{D}{2}k}\alpha_{\Omega k}^{\text{(II)}}, &\beta_{\Omega k}^{\text{(II)}}& \to e^{-i \frac{D}{2}k}\beta_{\Omega k}^{\text{(II)}}.
\end{align}
We can now compute the noise matrix characterizing the channel for $D\neq 0$. First, we observe that the diagonal blocks of $N$ remain unchanged as compared to the $D=0$ case, i. e. they are given by the expressions \eqref{Ns}. This is because these diagonal blocks correspond to local measurements in region I and region II, and as such they both remain invariant under translations of these regions due to the translational invariance of the Minkowski vacuum. Therefore we only need to evaluate the off-diagonal blocks of the matrix $N$, which carry information about correlations in the noise.

Following the method introduced in Appendix~\ref{D0} let us evaluate the output of the channel in Eq.~\eqref{GC} acting on the Minkowski vacuum $\sigma^{(d)}_\text{vac}$. For that we employ the definition of covariance matrix in Eq.~\eqref{CM}, together with the relations in Eq.~\eqref{detrinbogo}. We need to compute the upper right block of the covariance matrix which is equal to the transpose of the lower left block due to the symmetries of the covariance matrix. We find:
\begin{widetext}
\begin{align}\label{sigma13dp}
\left(\sigma^{(d)}_{\text{vac}}\right)_{13} &=
{}_\text{M}\bra{0} \left( \hat{d}_\text{I}+\hat{d}\d_\text{I} \right)
\left( \hat{d}_\text{II}+\hat{d}\d_\text{II} \right)\ket{0}_{\text{M}} =2\,\Re\,\,{}_\text{M}\bra{0}\hat{d}_{\text{I}}\hat{d}_{\text{II}}+\hat{d}_{\text{I}}\hat{d}_{\text{II}}\d\ket{0}_{\text{M}} 
\nonumber\\
&=
2\,\Re\int\!\!\!\!\int\mathrm{d}\Omega\,\mathrm{d}\Xi\,(\psi_\text{I},\wi[\Omega])\left[ (\psi_\text{II},\wii[\Xi])\,{}_\text{M}\bra{0}\bi[\Omega]\bii[\Xi]\ket{0}_{\text{M}}+
 (\psi_\text{II},\wii[\Xi])^\star\, {}_\text{M}\bra{0}\bi[\Omega]\bii[\Xi]\d\ket{0}_{\text{M}}\right].
\end{align}
After first substituting relations \eqref{rinmind0} and \eqref{modifiedbogos} into the above equation and then substituting for the Minkowski-Rindler Bogolyubov coefficients $\alpha^{\text{(II)}}_{\Omega k}$, $\beta^{\text{(I)}}_{\Omega k}$ and $\beta^{\text{(II)}}_{\Omega k}$ in terms of $\alpha^{\text{(I)}}_{\Omega k}$ using Eq.~\eqref{intermsofalpha} we get: 
\begin{align}
\left(\sigma^{(d)}_{\text{vac}}\right)_{13}&=2\,\Re\!\int\!\!\!\!\int\!\!\!\!\int\mathrm{d}\Omega\,\mathrm{d}\Xi\,\mathrm{d}k\,(\psi_\text{I},\wi[\Omega])\left[-(\psi_\text{II},\wii[\Xi])\,{\alpha^\text{(I)}_{\Omega k}}^\star{\beta^\text{(II)}_{\Xi k}}^{\star}+(\psi_\text{II},\wii[\Xi])^\star\,{\alpha^\text{(I)}_{\Omega k}}^\star{\alpha^\text{(II)}_{\Xi k}}\right]e^{iDk}\label{sigma13beforesub}\\
&=2\,\Re\!\int\!\!\!\!\int\!\!\!\!\int\mathrm{d}\Omega\,\mathrm{d}\Xi\,\mathrm{d}k\,(\psi_\text{I},\wi[\Omega])\left[(\psi_\text{II},\wii[\Xi])\,{\alpha^\text{(I)}_{\Omega k}}^{\star}{\alpha^\text{(I)}_{\Xi k}}e^{-\frac{\pi\Xi}{a}}+(\psi_\text{II},\wii[\Xi])^\star\,{\alpha^\text{(I)}_{\Omega k}}^\star{\alpha^\text{(I)}_{\Xi k}}^\star\right]e^{iDk}.\label{sigma13aftersub}
\end{align}
\end{widetext}
In order to proceed with the integration over $k$ we have to evaluate the following two integrals:

\begin{align}
I_-&=\int \mathrm{d}k {\alpha^\text{(I)}_{\Omega k}}^{\star}{\alpha^\text{(I)}_{\Xi k}}e^{iDk},\\
I_+&=\int \mathrm{d}k {\alpha^\text{(I)}_{\Omega k}}^\star{\alpha^\text{(I)}_{\Xi k}}^\star e^{iDk}.
\end{align}
We rewrite them using the explicit form of coefficients $\alpha^\text{(I)}_{\Omega k}$, as given in Eq.~(\ref{intermsofalpha}), to obtain:
\begin{align}
I_\pm&=\frac{e^{\frac{\pi(\Omega+\Xi)}{2a}}}{4\pi a\sqrt{\sinh\frac{\pi\Omega}{a}\sinh\frac{\pi\Xi}{a}}}\int\frac{\mathrm{d}k}{\omega_k}\left(\frac{\omega_k+k}{\omega_k-k}\right)^{i\frac{\Omega\pm\Xi}{2a}}e^{iDk}.
\end{align}
Let us introduce $\theta_\pm\equiv\frac{\Omega\pm\Xi}{a}$ and $\Delta\equiv mD$. We continue by changing the integration variable to $x = \text{asinh} \frac{k}{m}$, and treating the integral as a distribution, to get:
\begin{align}
I_\pm &\frac{4\pi a\sqrt{\sinh\frac{\pi\Omega}{a}\sinh\frac{\pi\Xi}{a}}}{e^{\frac{\pi(\Omega+\Xi)}{2a}}}=\int_{-\infty}^{\infty}\mathrm{d}x\,e^{i\left(\Delta\sinh x+\theta_\pm x\right)}=\nonumber\\
&=\int_{-\infty}^{\infty}\mathrm{d}x\,\cos(\Delta\sinh x + \theta_\pm x)=\nonumber\\
&=2\int_{0}^{\infty}\mathrm{d}x\,\cos\left(|\Delta|\sinh x\right)\cos(\theta_\pm x)\nonumber\\
&-2 \frac{\Delta}{|\Delta|}\int_{0}^{\infty}\mathrm{d}x\, \sin\left(|\Delta|\sinh x\right)\sin (\theta_\pm x).
\end{align}
The above integrals are of the form that is proportional to the following integral representation of the modified Bessel function \cite{NIST}:
\begin{align}
K_{i\nu}(\delta)&=\frac{1}{\cosh\frac{\pi\nu}{2}}\int_0^\infty \mathrm{d}x \cos(\delta\sinh x)\cos(\nu x)\nonumber\\
&=\frac{1}{\sinh\frac{\pi\nu}{2}}\int_0^\infty \mathrm{d}x \sin(\delta\sinh x)\sin(\nu x),
\end{align}
valid for $\delta >0$. Implementing the above identities into $I_\pm$ gives us:
\begin{align}
\label{ipmresult}
I_\pm&=\frac{ e^{\frac{\pi(\Omega+\Xi)}{2a}}}{2\pi a\sqrt{\sinh\frac{\pi\Omega}{a}\sinh\frac{\pi\Xi}{a}}} e^{-\frac{D}{|D|}\frac{\pi\theta_\pm}{2}}K_{i\theta_\pm}(|\Delta|).
\end{align}
This result can be used in Eq.~\eqref{sigma13aftersub}, to obtain the following value of the covariance matrix element $\left(\sigma^{(d)}_{\text{vac}}\right)_{13}$:
\begin{widetext}
\begin{align}\label{sameforDneg}
\left(\sigma^{(d)}_{\text{vac}}\right)_{13}=&\frac{1}{\pi a}\Re\!\int\!\!\!\!\int\mathrm{d}\Omega\mathrm{d}\Xi\frac{(\psi_\text{I},\wi[\Omega])}{\sqrt{\sinh\left(\frac{\pi\Omega}{a}\right)\sinh\left(\frac{\pi\Xi}{a}\right)}}\times\nonumber\\
&\times\left[e^{\frac{\pi(\Omega-\Xi)}{2a}(1-\frac{D}{|D|})}(\psi_\text{II},\wii[\Xi])K_{i(\frac{\Omega-\Xi}{a})}(m|D|)+e^{\frac{\pi(\Omega+\Xi)}{2a}(1-\frac{D}{|D|})}(\psi_\text{II},\wii[\Xi])^\star K_{i(\frac{\Omega+\Xi}{a})}(m|D|)\right].
\end{align} 
\end{widetext}
We compute the remaining elements of the upper right block of the covariance matrix by the same method used to calculate $\left(\sigma^{(d)}_{\text{vac}}\right)_{13}$. As a result we find that the noise matrix $N$ for $D\neq 0$ is still given by Eq.~\eqref{noise} with the expressions for $N_{\text{I}}$ and $N_{\text{II}}$ given by Eqs.~\eqref{N1} and \eqref{N2} while $N^\pm_{\text{I,II}}(D)$ takes the generalized form \eqref{N12forD}.

\subsection{When accelerations are in the same direction \label{parallelacc}}

Our calculation of the Gaussian channel proceeds in a similar manner as the $D\neq 0$ case. First of all, in this scenario the matrix $M$ given by Eq.~\eqref{M} and the diagonal elements of the noise matrix $N$ given by Eqs.~\eqref{N1} and \eqref{N2} are the same as in the previous scenarios. The only change is in the off-diagonal elements of the noise matrix given by $N^\pm_{\text{I,II}}$. In order to compute $N^\pm_{\text{I,II}}$ we first give the necessary  modifications of the Minkowski-Rindler Bogolyubov coefficients. Similar to the previous case we have:
\begin{align}\label{LRbogo}
\alpha_{\Omega k}^{\text{(I)}}& \longrightarrow e^{-i \frac{D}{2}k}\alpha_{\Omega k}^{\text{(I)}},
\nonumber\\
\beta_{\Omega k}^{\text{(I)}}&  \longrightarrow e^{i \frac{D}{2}k}\beta_{\Omega k}^{\text{(I)}}
= -\,e^{i \frac{D}{2}k} e^{-\frac{\pi \Omega}{a}} \alpha_{\Omega k}^{\text{(I)}},
\nonumber\\
\alpha_{\Omega k}^{\text{(II)}}& \longrightarrow e^{i \frac{D}{2} k}\alpha_{\Omega k}^{\text{(I)}},
\nonumber\\
\beta_{\Omega k}^{\text{(II)}}&  \longrightarrow e^{-i \frac{D}{2}k}\beta_{\Omega k}^{\text{(I)}}
= -\,e^{-i \frac{D}{2}k} e^{-\frac{\pi \Omega}{a}} \alpha_{\Omega k}^{\text{(I)}}.
\end{align}  
Substituting these Bogolyubov coefficients in \eqref{sigma13beforesub} and integrating over $k$ with the aid of the result of Eq.~\eqref{ipmresult} we get Eq.~\eqref{N12parallelfull}.

\section{Asymptotic properties of the modified Bessel function\label{N12limits}}
In this section, we study the limiting behavior of the modified Bessel function that is useful in finding the asymptotic behavior of the off-diagonal blocks of the noise matrix $N$ as $D\to0$. 

Let us begin by invoking the asymptotic expression for the modified Bessel function for small positive arguments $\epsilon$ \cite{Crispino2008}:
\begin{align}\label{MBK}
K_{i\nu}(2\epsilon)&\approx \frac{i\pi}{2\sinh(\pi\nu)}\left[\frac{\epsilon^{i\nu}}{\Gamma(1+i\nu)}-\frac{\epsilon^{-i\nu}}{\Gamma(1-i\nu)} \right],
\end{align}
where $\Gamma(z)$ is the Euler's Gamma function which  is defined as $\Gamma(z)=\int_{0}^{\infty}\mathrm{d}t\,e^{-t} t^{z-1}$, for complex numbers $z$ such that $\Re\,z>0$. When $\Re\,z\leq 0$, it is defined by analytic continuation and it has simple poles at nonpositive integer arguments. 

Using the identity $\Gamma(z)\Gamma(1-z)=\frac{\pi}{\sin(\pi z)}$, we can rewrite the asymptotic form  of $K_{i\nu}(2\epsilon)$, given in Eq.~\eqref{MBK}, as:
\begin{align}
K_{i\nu}(2\epsilon)&\approx \Re \left[\epsilon^{i\nu}\Gamma (-i\nu)\right].\label{Kapprox}
\end{align}
Note that $\Im\,\Gamma(ix)\propto-\frac{1}{x}$, which is divergent at $x=0$, can only be treated as a distribution, but $\Re\,\Gamma(ix)$ is regular at $x=0$. Let us substitute $\epsilon = e^{-\lambda}$ in Eq.~(\ref{Kapprox}) and look for the limit $\lambda\to\infty$:
\begin{align}\label{BKPV}
K_{i\nu}(2\epsilon)&\approx \cos(\lambda\nu)\,\Re\,\Gamma(-i\nu)+\sin(\lambda\nu)\,\Im\,\Gamma(-i\nu).
\end{align}
We can now take the limit of $\lambda\to\infty$. In doing so we will use the \textsl{Riemann-Lebesgue lemma}\footnote{Let $f$ be a Riemann-integrable function defined on an interval $a\leq x\leq b$ of the real line, the Riemann-Lebesgue lemma says $\lim_{|\alpha|\rightarrow\infty}\int_{a}^{b}f(x)\sin(\alpha x)\mathrm{d}x=\lim_{|\alpha|\rightarrow\infty}\int_{a}^{b}f(x)\cos(\alpha x)\mathrm{d}x=0$.}. According to it the limit of the first term in Eq.~\eqref{BKPV} vanishes. The limit of the second term can only be given in a distributional sense and it is non-zero only at the pole of the Gamma function at $\nu=0$:
\begin{align}
\lim_{\epsilon\to 0^+} K_{i\nu}(2\epsilon) &= \lim_{\lambda\to\infty }\sin(\lambda\nu)\, \Im\,\Gamma(-i\nu) = \lim_{\lambda\to\infty }\frac{\sin(\lambda\nu)}{\nu} \nonumber \\
&= \pi \delta(\nu),
\end{align}
where we have used one of representations of the Dirac delta.

\end{document}